\pdfoutput=1

\documentclass[aps,prd,twocolumn,superscriptaddress,notitlepage]{revtex4-1}

\usepackage{amssymb}
\usepackage{amsmath}
\usepackage{subfigure}
\usepackage[figuresright]{rotating}
\usepackage{multirow,tabularx}
\begin{document}

\title{Rev{\TeX} Author List for 20131121}

\affiliation{III. Physikalisches Institut, RWTH Aachen University, D-52056 Aachen, Germany}
\affiliation{School of Chemistry \& Physics, University of Adelaide, Adelaide SA, 5005 Australia}
\affiliation{Dept.~of Physics and Astronomy, University of Alaska Anchorage, 3211 Providence Dr., Anchorage, AK 99508, USA}
\affiliation{CTSPS, Clark-Atlanta University, Atlanta, GA 30314, USA}
\affiliation{School of Physics and Center for Relativistic Astrophysics, Georgia Institute of Technology, Atlanta, GA 30332, USA}
\affiliation{Dept.~of Physics, Southern University, Baton Rouge, LA 70813, USA}
\affiliation{Dept.~of Physics, University of California, Berkeley, CA 94720, USA}
\affiliation{Lawrence Berkeley National Laboratory, Berkeley, CA 94720, USA}
\affiliation{Institut f\"ur Physik, Humboldt-Universit\"at zu Berlin, D-12489 Berlin, Germany}
\affiliation{Fakult\"at f\"ur Physik \& Astronomie, Ruhr-Universit\"at Bochum, D-44780 Bochum, Germany}
\affiliation{Physikalisches Institut, Universit\"at Bonn, Nussallee 12, D-53115 Bonn, Germany}
\affiliation{Universit\'e Libre de Bruxelles, Science Faculty CP230, B-1050 Brussels, Belgium}
\affiliation{Vrije Universiteit Brussel, Dienst ELEM, B-1050 Brussels, Belgium}
\affiliation{Dept.~of Physics, Chiba University, Chiba 263-8522, Japan}
\affiliation{Dept.~of Physics and Astronomy, University of Canterbury, Private Bag 4800, Christchurch, New Zealand}
\affiliation{Dept.~of Physics, University of Maryland, College Park, MD 20742, USA}
\affiliation{Dept.~of Physics and Center for Cosmology and Astro-Particle Physics, Ohio State University, Columbus, OH 43210, USA}
\affiliation{Dept.~of Astronomy, Ohio State University, Columbus, OH 43210, USA}
\affiliation{Niels Bohr Institute, University of Copenhagen, DK-2100 Copenhagen, Denmark}
\affiliation{Dept.~of Physics, TU Dortmund University, D-44221 Dortmund, Germany}
\affiliation{Dept.~of Physics, University of Alberta, Edmonton, Alberta, Canada T6G 2E1}
\affiliation{Erlangen Centre for Astroparticle Physics, Friedrich-Alexander-Universit\"at Erlangen-N\"urnberg, D-91058 Erlangen, Germany}
\affiliation{D\'epartement de physique nucl\'eaire et corpusculaire, Universit\'e de Gen\`eve, CH-1211 Gen\`eve, Switzerland}
\affiliation{Dept.~of Physics and Astronomy, University of Gent, B-9000 Gent, Belgium}
\affiliation{Dept.~of Physics and Astronomy, University of California, Irvine, CA 92697, USA}
\affiliation{Laboratory for High Energy Physics, \'Ecole Polytechnique F\'ed\'erale, CH-1015 Lausanne, Switzerland}
\affiliation{Dept.~of Physics and Astronomy, University of Kansas, Lawrence, KS 66045, USA}
\affiliation{Dept.~of Astronomy, University of Wisconsin, Madison, WI 53706, USA}
\affiliation{Dept.~of Physics and Wisconsin IceCube Particle Astrophysics Center, University of Wisconsin, Madison, WI 53706, USA}
\affiliation{Institute of Physics, University of Mainz, Staudinger Weg 7, D-55099 Mainz, Germany}
\affiliation{Universit\'e de Mons, 7000 Mons, Belgium}
\affiliation{T.U. Munich, D-85748 Garching, Germany}
\affiliation{Bartol Research Institute and Dept.~of Physics and Astronomy, University of Delaware, Newark, DE 19716, USA}
\affiliation{Dept.~of Physics, University of Oxford, 1 Keble Road, Oxford OX1 3NP, UK}
\affiliation{Dept.~of Physics, University of Wisconsin, River Falls, WI 54022, USA}
\affiliation{Oskar Klein Centre and Dept.~of Physics, Stockholm University, SE-10691 Stockholm, Sweden}
\affiliation{Dept.~of Physics and Astronomy, Stony Brook University, Stony Brook, NY 11794-3800, USA}
\affiliation{Dept.~of Physics, Sungkyunkwan University, Suwon 440-746, Korea}
\affiliation{Dept.~of Physics, University of Toronto, Toronto, Ontario, Canada, M5S 1A7}
\affiliation{Dept.~of Physics and Astronomy, University of Alabama, Tuscaloosa, AL 35487, USA}
\affiliation{Dept.~of Astronomy and Astrophysics, Pennsylvania State University, University Park, PA 16802, USA}
\affiliation{Dept.~of Physics, Pennsylvania State University, University Park, PA 16802, USA}
\affiliation{Dept.~of Physics and Astronomy, Uppsala University, Box 516, S-75120 Uppsala, Sweden}
\affiliation{Dept.~of Physics, University of Wuppertal, D-42119 Wuppertal, Germany}
\affiliation{DESY, D-15735 Zeuthen, Germany}

\author{M.~G.~Aartsen}
\affiliation{School of Chemistry \& Physics, University of Adelaide, Adelaide SA, 5005 Australia}
\author{R.~Abbasi}
\affiliation{Dept.~of Physics and Wisconsin IceCube Particle Astrophysics Center, University of Wisconsin, Madison, WI 53706, USA}
\author{M.~Ackermann}
\affiliation{DESY, D-15735 Zeuthen, Germany}
\author{J.~Adams}
\affiliation{Dept.~of Physics and Astronomy, University of Canterbury, Private Bag 4800, Christchurch, New Zealand}
\author{J.~A.~Aguilar}
\affiliation{D\'epartement de physique nucl\'eaire et corpusculaire, Universit\'e de Gen\`eve, CH-1211 Gen\`eve, Switzerland}
\author{M.~Ahlers}
\affiliation{Dept.~of Physics and Wisconsin IceCube Particle Astrophysics Center, University of Wisconsin, Madison, WI 53706, USA}
\author{D.~Altmann}
\affiliation{Erlangen Centre for Astroparticle Physics, Friedrich-Alexander-Universit\"at Erlangen-N\"urnberg, D-91058 Erlangen, Germany}
\author{C.~Arguelles}
\affiliation{Dept.~of Physics and Wisconsin IceCube Particle Astrophysics Center, University of Wisconsin, Madison, WI 53706, USA}
\author{T.~C.~Arlen}
\affiliation{Dept.~of Physics, Pennsylvania State University, University Park, PA 16802, USA}
\author{J.~Auffenberg}
\affiliation{Dept.~of Physics and Wisconsin IceCube Particle Astrophysics Center, University of Wisconsin, Madison, WI 53706, USA}
% ...

\collaboration{IceCube Collaboration}
\noaffiliation

\title{Search for a diffuse flux of astrophysical muon neutrinos\\ with the IceCube 59-string configuration}

\begin{abstract}

%High-energy neutrinos are ideal messenger particles for the identification of cosmic-ray sources and their acceleration mechanisms, as neutrinos propagate unaffected through the Universe, maintaining directional and energetic information. 

%One of the open questions in astroparticle physics is the identification of cosmic-ray sources and their acceleration mechanisms.
%, which accelerate particles up to EeV energies and beyond.
% The study of charged cosmic rays has not yet revealed their origin, because charged particles are deflected by cosmic magnetic fields.
%, and extremely high-energy nuclei are attenuated by interactions with the cosmic microwave background. 
%High-energy neutrinos are ideal messenger particles for this search, as they propagate unaffected through the Universe, maintaining directional and energetic information. 
%The IceCube Neutrino Observatory was optimized for high-energy neutrino detection in the TeV to EeV energy range. 
%For the search for %high-energy astrophysical neutrinos with the IceCube Neutrino Observatory, a data sample of high-energy muon neutrinos with a very low background contamination of atmospheric muons was obtained with data taken between May 2009 and May 2010, when the IceCube detector was %still under construction, running in its 59-string configuration. 

A search for high-energy neutrinos was performed using data collected by the IceCube Neutrino Observatory from May 2009 to May 2010, when the array was running in its 59-string configuration. The data sample was optimized to contain muon neutrino induced events with a background contamination of atmospheric muons of less than 1\%. These data, which are dominated by atmospheric neutrinos, are analyzed with a global likelihood fit to search for possible contributions of prompt atmospheric and astrophysical neutrinos, neither of which have yet been identified. %The main signature of 
Such signals are %is the neutrino energy, which is 
expected to follow a harder energy spectrum than conventional atmospheric neutrinos. In addition, the zenith angle distribution differs for astrophysical and atmospheric signals. A global fit of the reconstructed energies and directions of observed events is performed, including possible neutrino flux contributions for an astrophysical signal and atmospheric backgrounds as well as systematic uncertainties of the experiment and theoretical predictions. The best fit yields an astrophysical signal flux for $\nu_\mu + \bar\nu_\mu $ 
of %at the level of 
$E^2 \cdot \Phi (E) = 0.25 \cdot 10^{-8}\,\textrm{GeV}\,\textrm{cm}^{-2}\,\textrm{s}^{-1}\,\textrm{sr}^{-1}$, and a zero %fitted 
prompt component. Although the sensitivity of this analysis for astrophysical neutrinos surpasses the Waxman and Bahcall upper bound, the experimental limit at 90\% confidence level is a factor of $1.5$ above at a flux of $E^2 \cdot \Phi (E) = 1.44 \cdot 10^{-8}\,\textrm{GeV}\,\textrm{cm}^{-2}\,\textrm{s}^{-1}\,\textrm{sr}^{-1}$.
\end{abstract}

\maketitle

\section{Introduction}
\label{intro}

High-energy neutrinos are believed to be ideal cosmic messenger particles in order to discover the enigmatic sources of high-energy cosmic rays. They are generated from the weak decay of charged mesons, in particular pions and kaons produced in hadronic interactions in, or close to, the sources. In generic scenarios \cite{Gaisser:1994yf, Learned:2000sw, Becker:2007sv} these neutrinos are expected to exhibit the same hard energy spectrum as the accelerated parent particles, yielding a typical differential spectrum $\Phi (E) \propto E^{-2}$.

%% A promising scenario for the production of high-energy cosmic rays is provided by the Fermi model of hadronic acceleration \cite{fermi1949origin}. It assumes the acceleration of charged particles in an iterative process via electromagnetic scattering of the charged particles with ionized gas clouds within the magnetic field of astrophysical objects. The process becomes efficient through the repeated crossing of shock fronts between media moving with different velocities. This stochastic process leads on average to an energy gain, accelerating particles up to energies constrained by the magnetic field strength and dimension of the object. These constraints indicate that cosmic rays with energies below the ankle are most likely of galactic origin, with the knee as a feature of transistion from lighter to heavier elements and an extragalactic origin above the ankle. Cosmic rays above the ankle are accelerated dominantly by extragalactic objects.\\ 

%% In the presence of a matter or radiation target inside an astrophysical object, charged particles can interact hadronically and produce mesons, in particular pions and kaons, which further decay into neutrinos and gamma rays. These neutrinos exhibit the same energy spectrum as the charged particles, which is predicted by the Fermi model to be $d\Phi/dE \propto E^{-2}$. While charged particles are subject to energy losses during propagation, the neutrino spectrum retains its energy dependence when it arrives at Earth.

To date, no cosmic high-energy neutrino sources have been found \cite{Abbasi:2010rd}. This motivates the complementary approach of a search for a diffuse flux of astrophysical neutrinos \cite{Barwick:2010}. A cumulative flux is composed of the integrated flux of all neutrino sources and could be detected even if the individual source fluxes are below the detection threshold, as long as the source population is large. Such a scenario is in particular imaginable for extragalactic sources, e.g.~Active Galactic Nuclei, which are among the candidate sources of ultra high-energy cosmic rays and could produce a detectable neutrino signal in the energy region between 10\,TeV and 10\,PeV \cite{Gaisser:1994yf,Learned:2000sw}.

\begin{figure*}
  \begin{center}
  \includegraphics[width=0.48\textwidth]{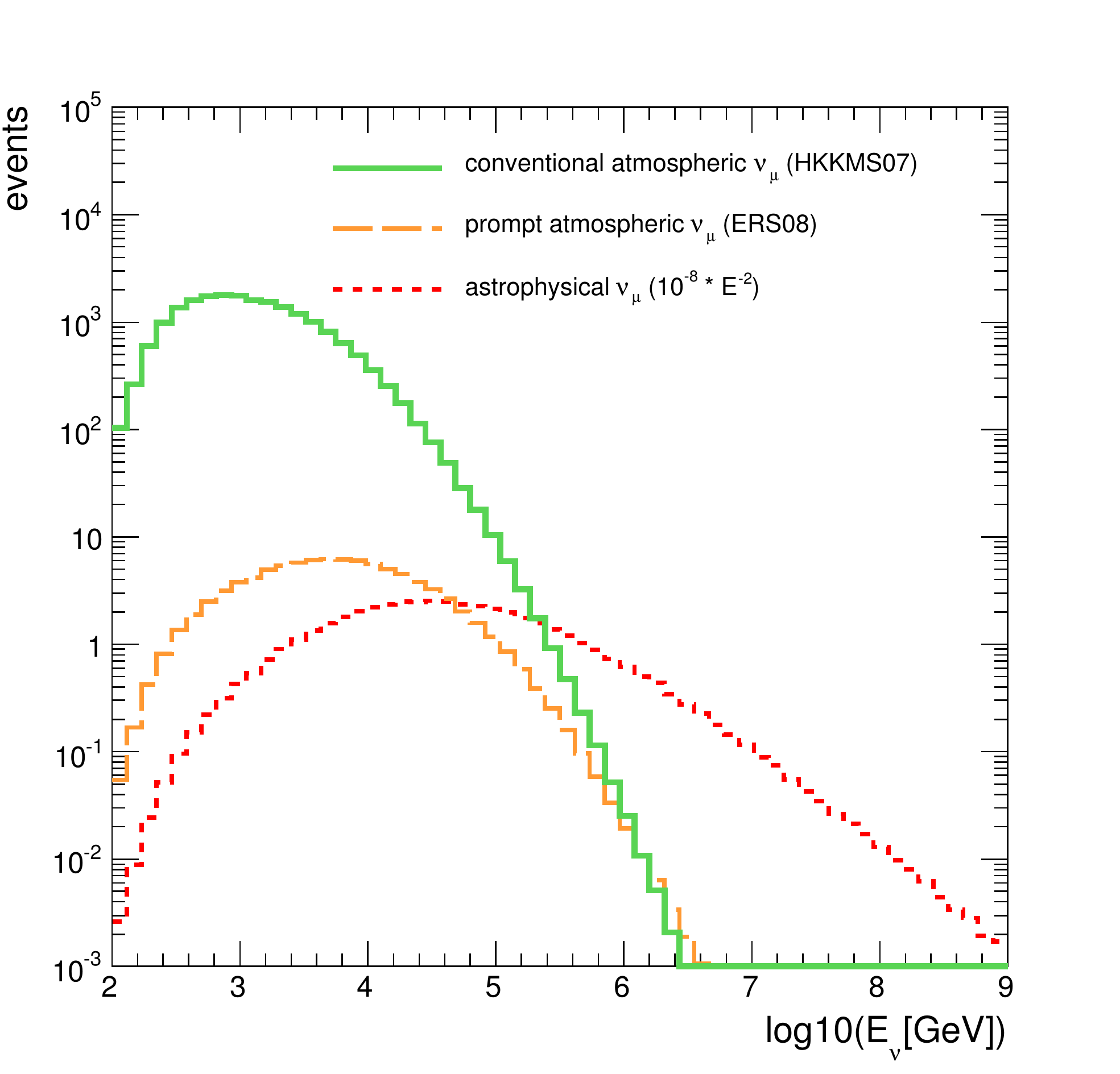}
  \includegraphics[width=0.48\textwidth]{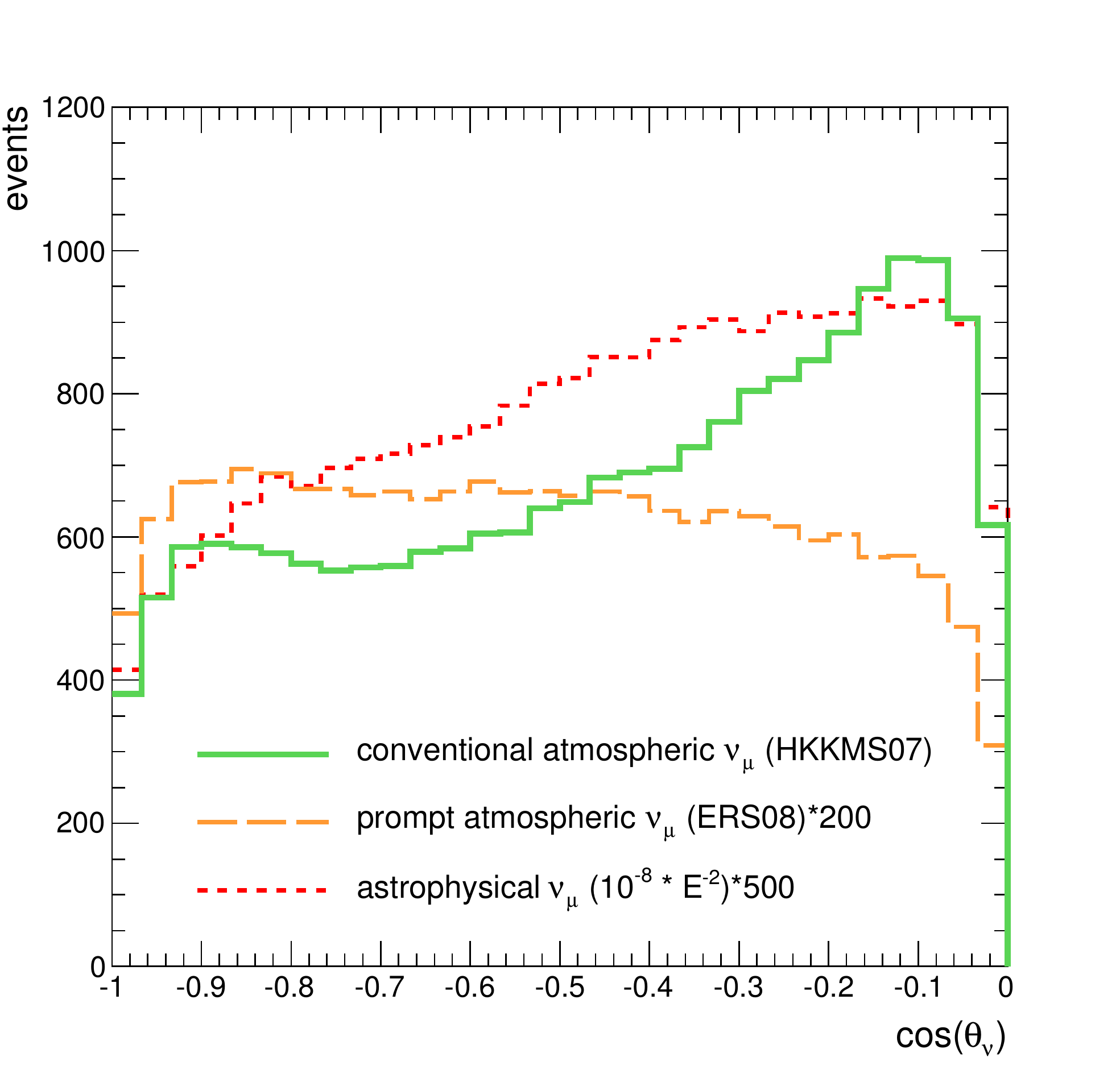}
  \end{center}
  \caption{Distribution of primary neutrino energies and zenith angles for conventional atmospheric \cite{Honda:2006qj, Gaisser:2012zz}, prompt atmospheric \cite{Enberg:2008te} and astrophysical $\nu_{\mu} + \overline \nu_{\mu}$ expected in the runtime of $348.1$ days, folded with the detection efficiency of this analysis. Note, the flux normalization of the latter two are multiplied by factors of 200 and 500 for better visibility in the right figure.}
  \label{img:TrueEnergyAndZenith}
\end{figure*}

The IceCube Neutrino Observatory is sensitive to diffuse fluxes of high-energy neutrinos of different flavors \cite{Aartsen:2013bka, eike, hese, Abbasi:2011jx} and has recently reported evidence for high-energy extraterrestrial neutrinos \cite{hese}. The analysis presented here searches for a diffuse astrophysical neutrino signal of high-energy upward-going muon tracks with the 59-string configuration, as previously done with smaller detector configurations \cite{Abbasi:2011jx}.
% and is a follow-up to a similar analysis with the 40-string configuration \cite{Abbasi:2011jx}. 
These analyses are mostly sensitive to charged-current interactions of muon neutrinos in the energy regime from a few TeV to several tens of PeV. A small sensitivity to charged-current interactions of tau neutrinos remains via taus decaying into muons. The field of view of these analyses is restricted to upward-going neutrinos in order to reject the dominant background of atmospheric muons (see section \ref{eventselection}). 

The main background to this search is the flux of atmospheric neutrinos, which is produced in cosmic-ray interactions with the Earth's atmosphere. The conventional atmospheric muons and neutrinos are produced in the decay of charged pions and kaons. Their energy spectrum is about one power steeper than the spectrum of the parent cosmic rays at Earth, due to the energy dependent competition between meson decay and interaction in the atmosphere. It is a power law with a spectral index of typically $\gamma = 3.7$.
%Therefore, the conventional atmospheric neutrino spectrum is a power law with a spectral index of typically $\gamma = 3.7$.

An additional atmospheric component is the flux of prompt atmospheric neutrinos: such neutrinos are produced in the decay of heavier mesons containing a charm quark. The cross sections for their production are small and therefore their contribution is only relevant at higher energies, where the conventional component is supressed below this level \cite{Enberg:2008te,Martin:2003us,Bugaev:1989we}. These heavy mesons have such short lifetimes that they immediately decay, rather than interact, which causes prompt neutrinos to follow the energy spectrum of the parent cosmic rays. They are a background for astrophysical neutrino searches at high energies, and have not yet been experimentally identified. Theoretical predictions of absolute fluxes are highly uncertain, mainly due to uncertainties in the parton distribution functions at very small values of Bjorken$-x$, which cannot be measured by collider experiments.
%because of large uncertainties in the differential cross sections for charm production through gluon or parton fusion. These depend on the parton density distribution at high energies and very small values of the Bjorken scaling variable $x$, which cannot be measured by collider experiments.

The different energy spectra of astrophysical, prompt and conventional atmospheric neutrinos are the main criteria for distinguishing the different components in the neutrino data sample measured with IceCube. This is illustrated in Fig.~\ref{img:TrueEnergyAndZenith}, which shows the expected energy and zenith angle distributions for the neutrino event selection used here (see section \ref{eventselection}).

An additional criterion is the zenith angle dependence: conventional atmospheric neutrinos exhibit a characteristic distribution with a maximum at the horizon. The reason is the angle dependent pathlengths of their parent mesons in the atmosphere, which determine their probability to decay and produce neutrinos before reaching the detector. As mentioned above, the mesons which produce prompt atmospheric neutrinos decay immediately, and therefore the prompt neutrinos are almost isotropically distributed. Assuming an isotropic distribution of astrophysical sources, the observed zenith angle distribution of astrophysical events is modified by the detector angular acceptance and the energy dependent absorption probability of neutrinos inside the Earth, which increases with energy. The absorption effect is stronger for astrophysical neutrinos than for prompt and conventional atmospheric neutrinos, due to their harder energy spectrum.

The connection between cosmic rays and astrophysical neutrinos permits an estimation of an upper bound for such a diffuse neutrino flux. The normalization of the neutrino flux to the observed cosmic-ray flux under the assumption of optimistic parameters for the efficiency of hadronic neutrino production in optically thin sources without re-acceleration of decaying parent particles leads to an upper bound of $E_{\nu}^2 \cdot \Phi (E_{\nu}) \sim 10^{-8} \textrm{GeV}\,\textrm{cm}^{-2}\,\textrm{s}^{-1}\,\textrm{sr}^{-1}$ as calculated by Waxman and Bahcall \cite{Waxman:1998yy,Waxman:2011hr}. Other model predictions for diffuse neutrino fluxes are based on the observed photon flux at different wavelengths from different experiments and can be above or below this upper bound (see section \ref{results}). The analysis presented in this paper reaches a sensitivity below the Waxman-Bahcall upper bound.

This paper is organized as follows: Section \ref{eventselection} describes the IceCube detector and the selection of upward-going muon neutrino events. The likelihood fit, which was chosen as an analysis method, and the treatment of systematic uncertainties in this fit are explained in sections \ref{likelihood} and \ref{systematics}. Section \ref{results} presents and discusses the results and the conclusion in section \ref{conclusion} summarizes and compares to other analyses.

\section{IceCube detector and data selection}
\label{eventselection}

%\begin{figure}
%  \begin{center}
%this is vector but large  \includegraphics[width=0.35\textwidth]{ArrayWSeasons-crop.pdf}
% temporarily use the png version for smaller size
%  \includegraphics[width=0.35\textwidth]{ArrayWSeasons-crop.png}
%  \end{center}
%  \caption{Sketch of the IceCube Neutrino Observatory.}%
%  \label{img:IceCubeSketch}%
%\end{figure}

IceCube is a neutrino detector located at the geographic South Pole \cite{Achterberg:2006md}. In neutrino interactions with nuclei, secondary particles are produced, which travel faster than the speed of light in the Antarctic ice and threrefore emit Cherenkov light. These photons are detected by optical sensors deployed in the Antarctic ice. In the final detector configuration, the digital optical modules (DOMs) are arranged on 86 vertical strings of 60 sensors each spread over depths between 1450 and 2450\,m with vertical distances of 17\,m between sensors. Seventy-eight strings have a horizontal spacing of about 125\,m and span a hexagon of a surface area of roughly 1\,km$^2$.
% (see Fig.~\ref{img:IceCubeSketch}). 
A further eight strings together with the seven surrounding IceCube strings form the more densely instrumented central DeepCore detector\cite{Collaboration:2011ym}. The IceCube detector was completed in December 2010. The analysis presented here was performed with data taken between May 2009 and May 2010, when IceCube was still under construction and consisted of 59 strings.

\begin{figure}
  \includegraphics[width=\linewidth]{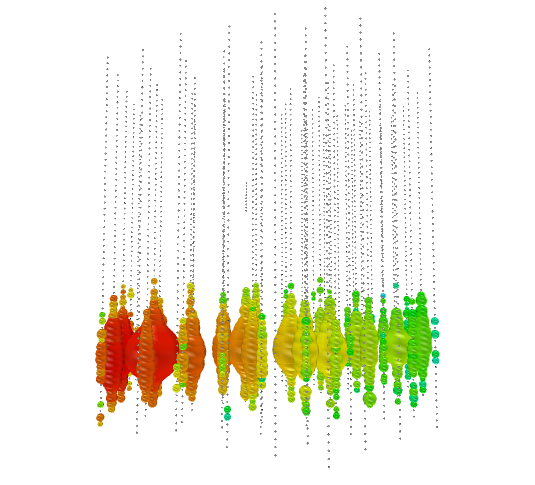}
  \caption{Event view of the highest-energy neutrino event observed in this analysis. The grey dots mark IceCube DOMs. DOMs hit by photons are shown in color. The color code indicates the photon arrival time with red colors marking early times and blue colors standing for late times. The radius of the DOMs correlates with the observed charge. The reconstructed zenith angle of this event is $91.2^{\circ}\pm 0.1^{\circ}$ and the reconstructed, truncated muon energy loss is $\log(dE/dx\,\textrm{[GeV/m]}) = 1.37$ within the detection volume. Assuming the best-fit energy spectrum from this analysis (see Fig.~\ref{img:EnergyResolution}), this event most likely  originated from a neutrino of energy 500TeV-1PeV, producing a muon that
  passed through the 
detector with an energy of about 400 TeV. }
  \label{img:Apollon}%
\end{figure}

The digital optical modules contain a photomultiplier tube (PMT) housed in a borosilicate glass pressure sphere. The PMT quantum efficiency as well as the transparency of the glass and the optical gel make the module most sensitive to wavelengths in the ultraviolet and blue regions \cite{Abbasi:2010vc}. This is optimal for the Cherenkov radiation filtered through Antarctic ice. If a trigger condition is fulfilled, the recorded waveforms are digitized and transferred to the surface. The quantities, which are extracted from the measured waveforms of each DOM, are the total number and arrival times of PMT photo-electron pulses, corresponding to the detected Cherenkov photons (see Fig.~\ref{img:Apollon}). This information is used for the reconstruction of the direction and estimation of the energy of the secondary particles, which is highly correlated to the initial neutrino direction and energy \cite{Ahrens:2003fg, Abbasi:2012wht, energypaper}.

The typical trigger condition for high-energy neutrino analyses in IceCube requires at least eight sensors recording light within a time window of $5\,\mu\textrm{s}$. The triggering sensors must be in a local coincidence with either of their neighboring or next-to-nearest neighboring  sensors. Most of the triggers come from atmospheric muons, which are produced in cosmic-ray air showers in the Earth atmosphere and are the main background in the search for neutrinos. With the 59-string configuration, the trigger rate for atmospheric muons was 1500\,Hz and initially outnumbered the rate of atmospheric neutrino-induced tracks by more than five orders of magnitude (see Tab.~\ref{tab:PassingEfficiencies}). A significant contribution to the atmospheric muon trigger rate comes from muons from coincident but independent air showers (coincident muons), which are particulary challenging to identify. The analysis requires a neutrino sample with a very low background contamination of atmospheric muons while retaining as many high-energy neutrino events as possible. It is optimized for the detection of through-going muons originating from muon neutrino charged-current interactions, which cause a track-like signature in the detector. The separation of neutrino-induced events from atmospheric muons is based on several steps: in the online-processing at the South Pole, potentially interesting events are selected and transmitted to the data-center in the North via satellite. During the offline-processing, more advanced reconstructions are performed and the data stream is further reduced through a pre-selection of highly energetic tracks. A high-purity muon neutrino sample is finally obtained through a series of quality cuts on reconstruction quality parameters. These steps are described in the following paragraphs.

At the South Pole, isolated noise pulses are excluded from the reconstruction. An online filter criterion optimized for track-like signatures reduces the data stream and selects high-energy muon candidate events. It requires a minimum amount of detected total charge and a minimum quality of a likelihood track reconstruction. The rejection of atmospheric muons takes advantage of the fact that muons are absorbed in matter, while neutrinos are able to traverse the Earth and are the only particles arriving at the detector from below.
Therefore, the filter criteria depend on the result of a fast first-guess angular reconstruction algorithm (Linefit) and are stronger for downward-going than for upward-going events. When transmitted to the North, atmospheric muons still dominate the neutrinos by a factor $10^4$.

During the offline-processing, further reconstructions, in particular an iterative likelihood fit including the number of detected photons (MPE likelihood), are performed. Before this reconstruction, the event's hit pattern is searched for subsets of causally connected pulses in order to remove remaining noise pulses and to identify pulses from coincident particles. For $\lesssim 50\%$ of the events, subsets of pulses are found which are not causally connected with the main cluster of pulses. These pulses
%deemed as noise, or as a subthreshold second particle, 
are ignored during that reconstruction. As a pre-selection of high-energy neutrino events, the field of view of the analysis is completely restricted to the upward-going region with zenith angles $\theta > 90^{\circ}$ (MPE fit). Additionally, a minimum number of hit sensors sufficiently close enough to the reconstructed track hypothesis to observe unscattered secondary photons is required.

The high-level event selection is developed through a comparison of Monte Carlo generated neutrino event signatures and atmospheric muon signatures. Neutrino events are generated and propagated through the Earth to a region surrounding the detector where their interactions in the rock and ice are simulated \cite{Gazizov:2004va}. Neutrino-induced muons are then tracked into and through the detector taking account of stochastic
and continuous energy losses \cite{Chirkin:2004hz}. Cherenkov light from charged particles is propagated to the optical modules \cite{Lundberg:2007mf} taking account of scattering and absorption in the ice \cite{ackermann2006optical, Aartsen:2013rt}. Finally, the generation of the signal as a function of time in the optical module is also simulated in detail. The background of atmospheric muons is simulated with the air shower simulation software CORSIKA \cite{Heck:1998vt} and from there on their simulation is passed through the same simulation chain as the neutrinos.

Generated neutrino events are re-weighted to a primary astrophysical or atmospheric neutrino spectrum of choice. In this analysis, the baseline model to describe the incoming flux of conventional atmospheric neutrinos is based on the model HKKMS07 \cite{Honda:2006qj}. The calculations in Refs.~\cite{Honda:2006qj, Barr:2004br} extend only to $E_\nu=10$~TeV. In previous IceCube analyses~\cite{Abbasi:2010ie, Abbasi:2011jx} these results have been extrapolated to higher energy by fitting a standard parameteriztion~\cite{GaisserTextbook} 
\begin{equation}
\begin{split}
&\Phi (E_{\nu}) \simeq \Phi_0 \cdot E_{\nu}^{-\gamma} \cdot \\
&\left( \frac{A_{\pi \nu}}{1 + B_{\pi \nu} E_{\nu} \cos( \theta^*)/\epsilon_{\pi}} + \frac{A_{K \nu}}{1 + B_{K \nu} E_{\nu} \cos( \theta^*)/\epsilon_{K}} \right)
\end{split}
\label{eq:gaisserneutrinoflux}
\end{equation}
to the published neutrino calculations below 10 TeV. In this equation, $\theta^*$ is the zenith angle where the neutrinos are produced, taking account of the curvature of the Earth~\cite{Lipari:1993hd}. The parameters $\Phi_0$, $A$ and $B$ are free fit parameters, the spectral index is $\gamma = 2.7$, and the critical energies are $\epsilon_{\pi} = 115\,$GeV and $\epsilon_{K} = 850\,$GeV. Such an extrapolation does not account for the knee in which the overall spectrum of the cosmic rays becomes steeper between 1 and 10 PeV.

%model to energies above 10\,TeV. The high-energy extrapolation has been improved compared to previous analyses \cite{Abbasi:2011jx} through the inclusion of realistic cosmic-ray flux parameterizations H3a \cite{Gaisser:2012zz} and poly-gonato \cite{hoerandel2003knee}, which take into account recent measurements of the cosmic-ray flux normalization, composition and the feature of the knee. This is discussed in section \ref{systematics}. The prompt atmospheric neutrino flux is estimated according to the prediction by Enberg \it{et al.} as a baseline model, which has also been modified at high-energies to take into account the cosmic-ray flux parameterizations of H3a and poly-gonato.

This analysis extends to PeV neutrino energies and therefore the steepening at the knee has to be accounted for. Since neutrino production occurs at the level of interactions of individual nucleons and mesons, a parameterization of the evolution of the elemental composition through the knee region is needed. Two different parameterizations, H3a of Ref.~\cite{Gaisser:2012zz} and a modified version of the poly-gonato parameterization~\cite{hoerandel2003knee} in which its galactic component is supplemented with an extragalactic component of the form of Ref.~\cite{Gaisser:2012zz}, are considered. The effect of the knee is implemented by folding the yield of neutrinos per primary nucleon with the primary spectrum of nucleons, as described in the Appendix. The prompt atmospheric neutrino flux is estimated according to the prediction by Enberg \textit{et al.}~(ERS08) \cite{Enberg:2008te} as a baseline model, which has also been modified at high-energies to take into account the cosmic-ray flux parameterizations of H3a and poly-gonato (see Appendix).

After the offline-processing and the pre-selection of high-energy upward-going tracks, the remaining backgrounds are mis-reconstructed events, often caused by coincident muons or muons passing outside the instrumented volume, which, despite being truly downward-going, are reconstructed as upward-going. Such background events are removed by selecting only upward-going events of high offline-reconstruction quality. This is accomplished through selecting on conditions on a number of parameters. The parameters are described in Ref.~\cite{Ahrens:2003fg, AnnePhd} and have their origin in five different ways of identifying events which are likely to have been poorly reconstructed. These parameters and the selection criteria
%, which have been found during the optimization of the event selection, 
are listed below and summarized in Tab.~\ref{tab:ListOfCuts}. 

\begin{table*}\index{typefaces!sizes}
  \footnotesize%
  \caption{List of event selection criteria and corresponding passing efficiencies. The passing efficiencies are given with respect to the previous step. The astrophysical neutrino flux is estimated assuming an $E^{-2}$ power law, and the conventional atmospheric neutrino flux is based on the prediction by Honda \textit{et al.}~(HKKMS07) \cite{Honda:2006qj} including the modification of the H3a cosmic-ray flux parameterization \cite{Gaisser:2012zz}.}
  \begin{center}
    \begin{tabular}{c|c||c|c|c}
      \hline
      group & selection criterion & \multicolumn{3}{c}{Passing efficiencies} \\
         & & atms. $\mu$ (coincident) & astrophysical & conv. atms. $\nu_{\mu}$\\
      \hline
      \multirow{2}{*}{1} & $\theta \textrm{ (MPE) } > 90^{\circ}$ & $95\%$ ($95\%$) & $98\%$ & $97\%$ \\
       & $\theta \textrm{ (Linefit) } > 90^{\circ}$ & $65\%$ ($57\%$) & $92\%$ & $91\%$ \\
      \hline
       & $rlogl < 11$ & $33\%$ ($42\%$) & $93\%$ & $75\%$ \\
      2 & $\sigma_{\textrm{\tiny{paraboloid}}} < 5^{\circ}$ & $19\%$ ($37\%$) & $77\%$ & $67\%$ \\
       & $ \psi \left(\textrm{Linefit, MPE}\right) < 15^{\circ}$ & $39\%$ ($49\%$) & $90\%$ & $92\%$ \\
      \hline
      \multirow{2}{*}{3} & $\log \left( \frac{\mathcal{L}_{\textrm{\tiny{SPE}}}}{\mathcal{L}_{\textrm{\tiny{Bayesian}}}}\right) > 29$ & $22\%$ ($30\%$) & $89\%$ & $67\%$ \\
       & $\min(\theta_{geo1}, \theta_{geo2}, \theta_{time1}, \theta_{time2}) > 80^{\circ}$ & $3\%$ ($2\%$) & $84\%$ & $67\%$ \\
      \hline
       & $N_{\textrm{\tiny{dir}}} > 6$ & $3\%$ ($2\%$) & $97\%$ & $94\%$ \\
      4 & $L_{\textrm{\tiny{dir}}} > 250\,\textrm{m}$ & $93\%$ ($91\%$) & $98\%$ & $97\%$ \\
       & $|S_{\textrm{\tiny{dir}}}| < 0.45$ & $30\%$ ($43\%$) & $97\%$ & $96\%$ \\
      \hline
      5 & $-450\,\textrm{m} < z_{\textrm{\tiny{COG}}} < 400\,\textrm{m}$ & $87\%$ ($96\%$) & $96\%$ & $97\%$ \\
      \hline
    \end{tabular}
  \end{center}
  \label{tab:ListOfCuts}
\end{table*}

%This is accomplished by a series of quality criteria, which are described in Ref.~\cite{Ahrens:2003fg, AnnePhd} and can be grouped into five different categories:

\begin{enumerate}
%\item Directional consistency requires that the two reconstruction algorithms MPE likelihood fit and Linefit converge to an upward-going track ($\theta > 90^{\circ} $) and have to agree within $\psi \le 15^{\circ}$ of each other.

\item The upward-going condition, $\theta > 90^{\circ}$, is required to be satisfied for the zenith angles found in the two angular reconstruction algorithms, MPE likelihood fit and Linefit.

\item A minimum track reconstruction quality is required based on the reduced negative log-likelihood at the minimum $rlogl = - \log \mathcal{L}/(N_{\textrm{\tiny{ch}}} - 5)$, where $N_{ch}$ is the number of hit sensors in the event. Additionally, the angular error estimation of the MPE likelihood fit $\sigma_{\textrm{\tiny{paraboloid}}}$ has to be smaller than $5^{\circ}$. Directional consistency between the two reconstruction algorithms, MPE likelihood fit and Linefit, is required through a condition that the difference between the zenith angles obtained from each algorithm $\psi$ satisfies  $\psi \leq 15 ^{\circ}$. 

\item The rejection of mis-reconstructed atmospheric muons is improved by a cut on the likelihood ratio of the reconstructed solution to a second reconstruction which is forced to a downward-going track and in which the likelihood is weighted with a Bayesian prior describing the probability that a downward-going muon is expected at that reconstructed zenith angle. In addition, individual reconstructions are performed on the hit pattern split in half based on geometry or time. All reconstructions of each split hit pattern have to fulfill $\theta > 80^{\circ}$.

\item In order to guarantee a matching between the track hypothesis and the measured hit pattern, a minimum number of direct hits $N_{\textrm{\tiny{dir}}}$, i.e. pulses that are recorded within a time window of $-15$ to $75\,$ns of the geometrically expected arrival time and therefore attributed to unscattered photons, is required. Further, the direct length $L_{\textrm{\tiny{dir}}}$, which is determined by the projection of the direct hits on the reconstructed track, has to exceed a certain minimum length. Additionally, these direct hits have to occur continuously along the reconstructed track, quantified by the smoothness variable $S_{\textrm{\tiny{dir}}}$.

\item Background events, which may pass above or below the detector and are very hard to reconstruct in direction and energy, are rejected by the
requirement that the position of the center of gravity of hit optical modules in the vertical direction ($z_{\textrm{\tiny{COG}}}$) is not at the top or bottom of the detector.

\end{enumerate}

The passing efficiencies are summarized in Tab.~\ref{tab:PassingEfficiencies}. The data selection was optimized keeping the signal region of the experimental data blind in order to avoid introducing a bias in the analysis. The signal region is defined as the 5\% of events with the highest reconstructed energy loss.

\begin{table*}\index{typefaces!sizes}
  \footnotesize%
  \caption{Measured and expected event rates in Hz for the IceCube 59-string data stream with a total livetime of 348.1 days. Atmospheric muon background expectations are based on CORSIKA simulation. Predictions for conventional atmospheric neutrinos are based on the prediction by Honda \textit{et al.}~(HKKMS07) \cite{Honda:2006qj} including the modification of the H3a cosmic-ray flux parameterization \cite{Gaisser:2012zz} and scaled to the best-fit nuisance parameters obtained later in this analysis (see section \ref{results}). The prompt atmospheric neutrino flux given in the table corresponds to the prediction ERS08 \cite{Enberg:2008te} and has also been modified based on the H3a parameterization.}
  \begin{center}
    \begin{tabular}{l|ccccc}
      \hline
         & experimental & atms. $\mu$ & astrophysical & conv. atms. $\nu_{\mu}$ & prompt atms. $\nu_{\mu}$\\
         & data & CORSIKA  & $\nu_{\mu}$ ($\nu_{\mu} + \nu_{\tau}$) & HKKMS07 + H3a & ERS08 + H3a\\
         & & total (coincident) &  $10^{-8}\,\mathrm{GeV\,cm^{-2}\,s^{-1}\,sr^{-1}} \cdot E^{-2}$ & best fit & + H3a\\
%         & & & $d\Phi/dE = 10^{-8} \cdot E^{-2}$ & best fit) & + H3a)\\
      \hline
      \hline
      %passing efficiency & $3.6\cdot 10^{-2}\%$ & $5.7\cdot 10^{-5}\%$ & $1.8\cdot 10^{-4}\%$ & $39.6\%$ & $15.8\%$\\
      %$\#$ at final level & $21943$ & $29.19$ & $21.20$ & $46.48$ & $19789.65$\\
      %$\#$ at trigger level & $XXX$ & $XXX$ & & & \\
      %$\#$ satellite transmitted & $XXX$ & $XXX (XXX)$ & $XXX (XXX)$ & $XXX$ & $XXX$\\
      %$\#$ at final level & $21943$ & $29$ ($21$) & $46$ ($50$) & $21620$ & $91$\\
      trigger level [Hz] & $1.5 \cdot 10^3$ & $1.4 \cdot 10^3$ & & $2.4 \cdot 10^{-2}$ & \\
      satellite transmitted [Hz] & $35.2$ & $30.2$ & & $8 \cdot 10^{-3}$ & \\
      %high-energy pre-selection & $2.1$ & $1.4$ ($0.2$) & $3.9 \cdot 10^{-6}\,$ & $4.5 \cdot 10^{-3}\,$ & $1.0 \cdot 10^{-5}\,$\\
      at final level [Hz] & $7.3 \cdot 10^{-4}\,$ & $9.6 \cdot 10^{-7}\,$ ($7.0 \cdot 10^{-7}\,$) & $1.5 \cdot 10^{-6}\,$ ($1.7 \cdot 10^{-6}\,$) & $7.2 \cdot 10^{-4}\,$ & $3.0 \cdot 10^{-6}\,$\\
      \hline
      \hline
      $\# \nu$ at final level in 348 days & $21943$ & $29$ ($21$) & $46$ ($50$) & $21844$ & $91$\\
      \hline
    \end{tabular}
  \end{center}
%The row \emph{high-energy pre-selection} corresponds to a pre-selection of upward-going high-energy track candidate events on which further time consuming reconstruction algorithms are applied.
%\Comment{(Comment)}{Numbers for atm, nu are inconsistent with annes appendix C.. need to double check}
%\Comment{(Comment)}{Line 3 and 4 are derived from latest MC, that was also used in the analysis. Line 2 is taken from the muon filter proposal, which means these numbers were estimated with very early simulation (probably containing nugen bug and others). A possibility is to estimate the numbers for line 2 from the passing efficiencies we know from the level 3 cuts (which are 7\% for atms. muons, 76\% for astrophysical nus and 74\% for atmospheric nus), for the upward-going region (which means another factor 2). However, these passing efficiencies (see appendix C, thesis) have also been estimated from not the latest MC version.}
  \label{tab:PassingEfficiencies}
\end{table*}

As listed in Tab.~\ref{tab:PassingEfficiencies}, the final experimental data sample consists of $21943$ events acquired within a total livetime of $348.1$ days. The sample is expected to be dominated by conventional atmospheric neutrinos with an expected number of 21844 events based on the HKKMS07 model, including the modification of the H3a cosmic-ray flux parameterization. The expected number of prompt neutrinos is $91$ for the model ERS08 modified to correspond to the H3a cosmic-ray flux parameterization. An astrophysical flux ($\nu_{\mu} + \nu_{\tau}$) at the level of the Waxman-Bahcall upper bound would correspond to about $50$ events in this data sample. The contamination of background from misreconstructed atmospheric muons in this event selection is estimated from simulations. These calculations find a neutrino purity of $99.85\% \pm 0.06\%\,\textrm{(stat.)}\,\pm 0.04\%\,\textrm{(sys.)}$, corresponding to a muon background of about $30$ events. As these muons are rather low in energy they are not included in the further analysis.

\begin{figure}
  \begin{center}
  \includegraphics[width=0.45\textwidth]{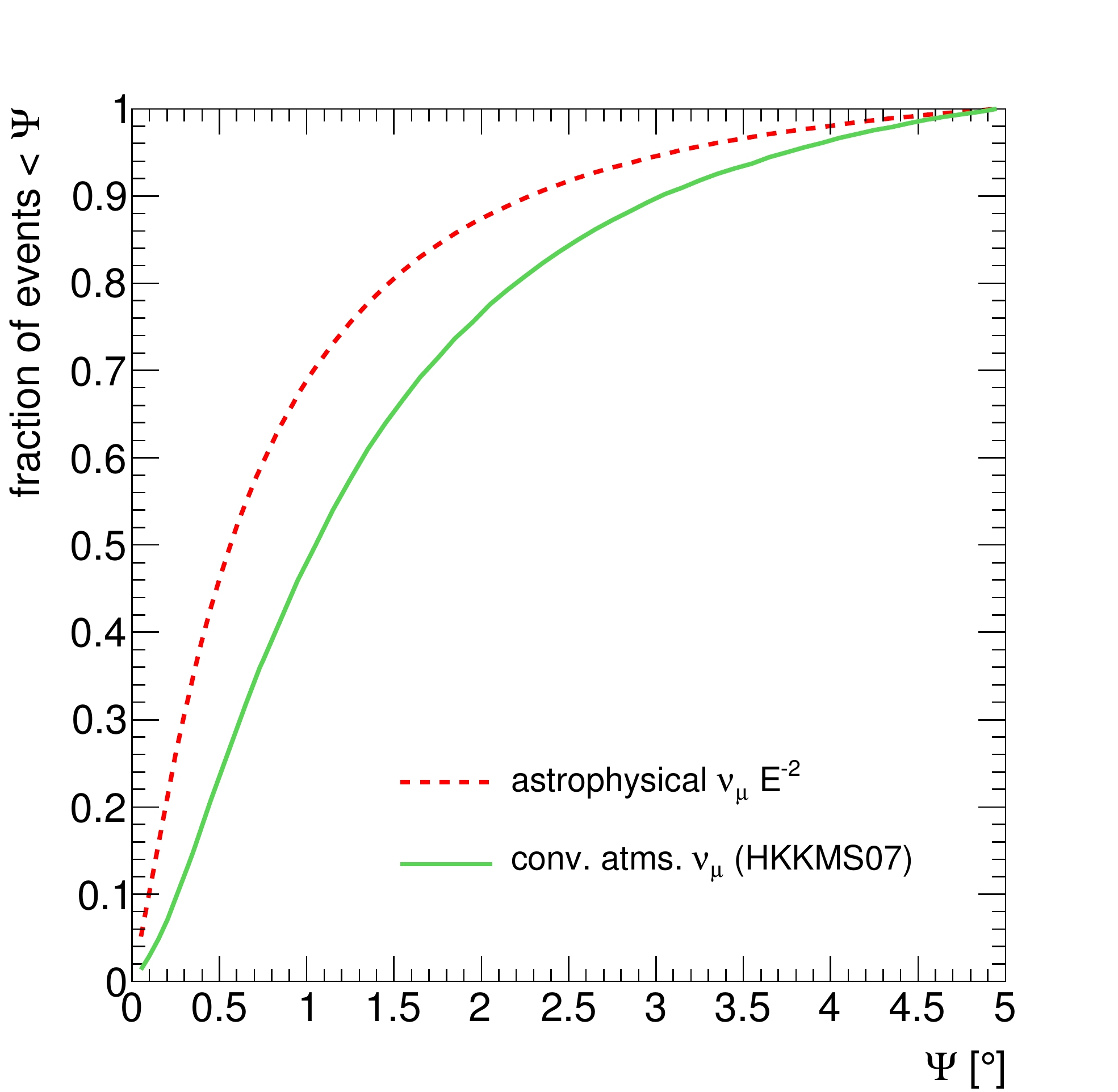}
  \end{center}
  \caption{The cumulative distribution of the angular resolution for astrophysical $E^{-2}$ and conventional atmospheric events reconstructed by the MPE fit \cite{Ahrens:2003fg} obtained from Monte Carlo simulation.}
%\Comment{(Comment)}{Left plot lable should be $\psi$ instead of $\Delta \psi$.}
  \label{img:ZenithResolution}
\end{figure}

\begin{figure*}
  \begin{center}
  \includegraphics[width=0.32\textwidth]{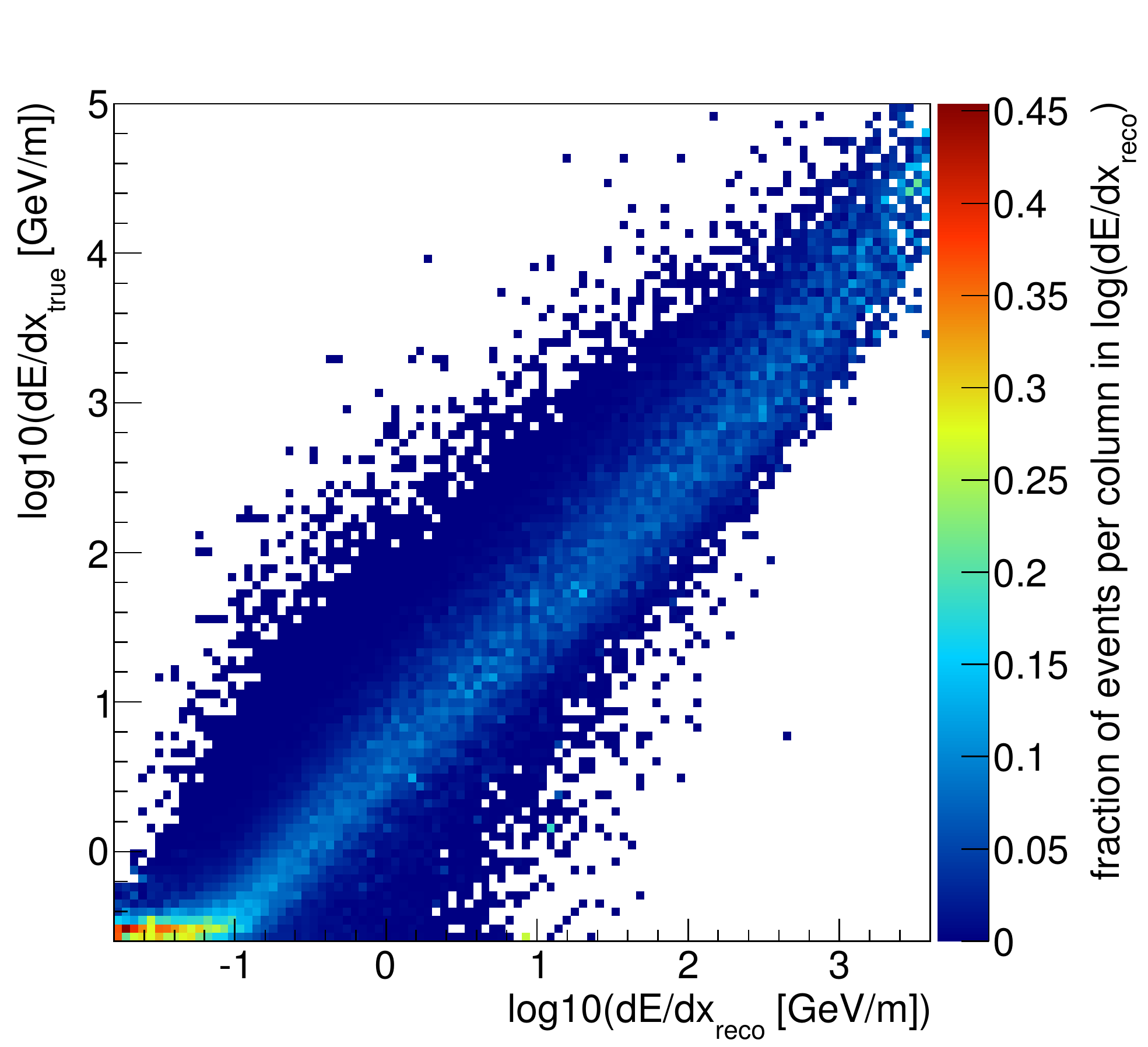}
  \includegraphics[width=0.32\textwidth]{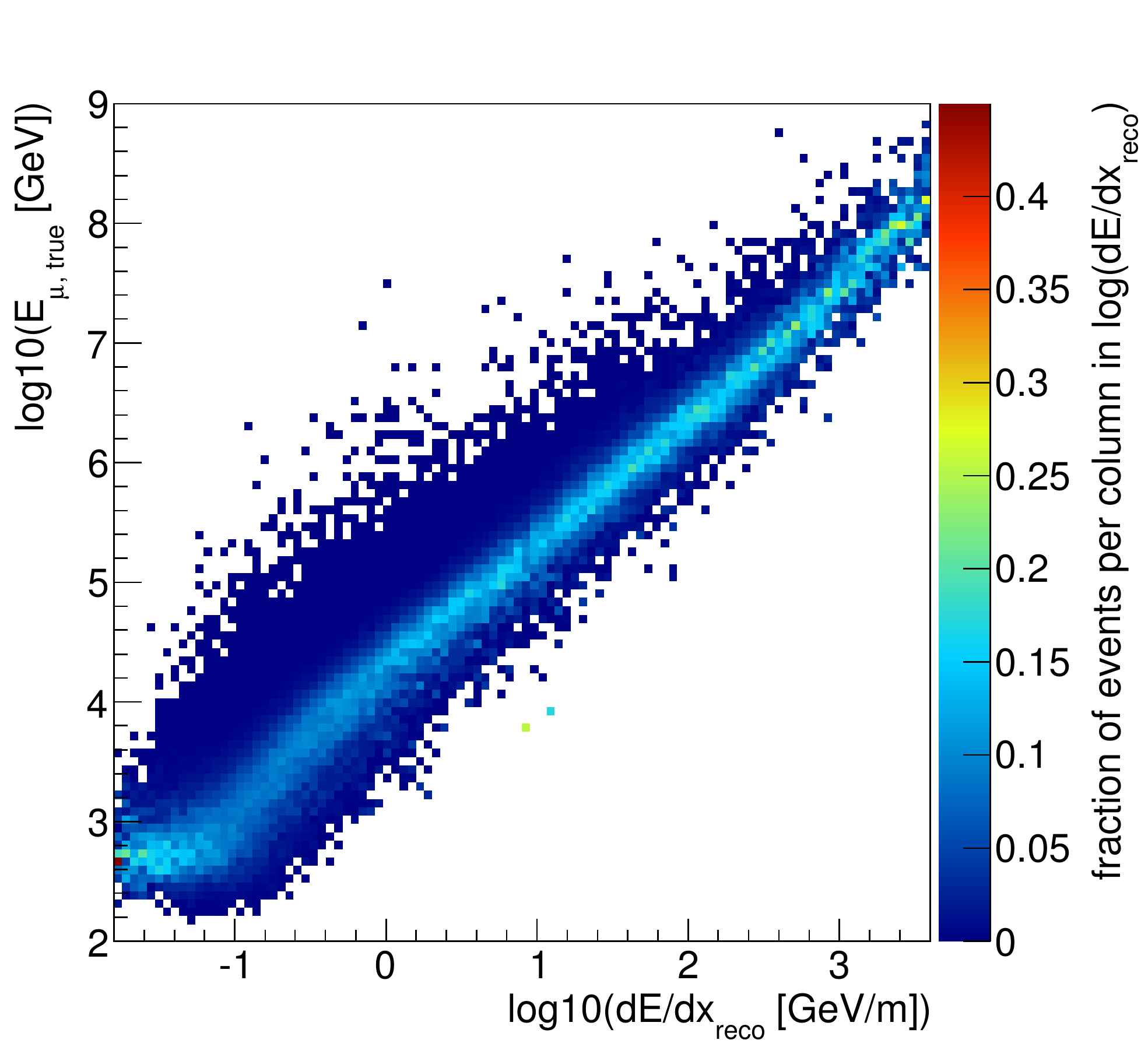}
  \includegraphics[width=0.32\textwidth]{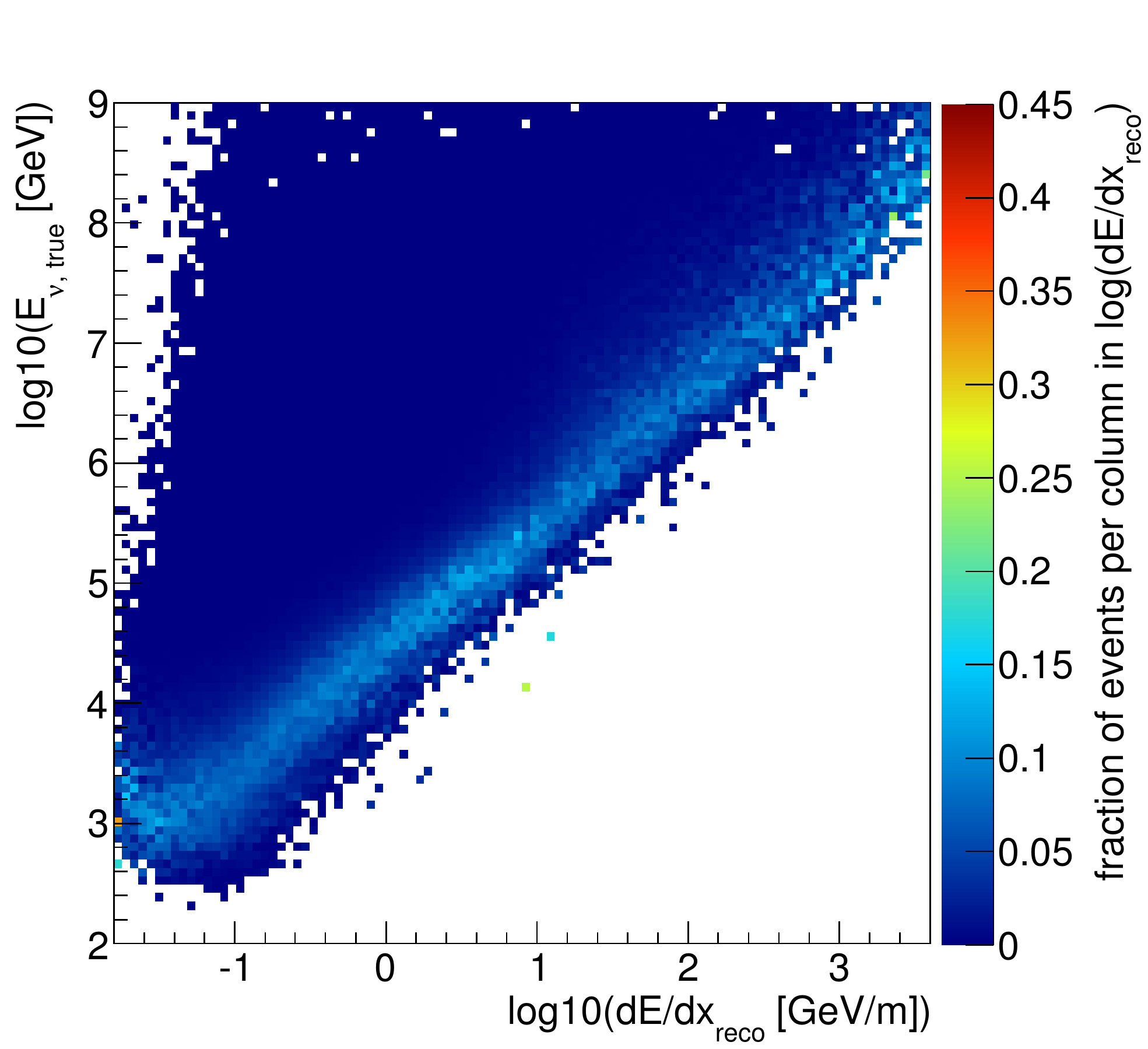}
  \end{center}
  \caption{Correlation between the truncated energy loss of the muon reconstructed with the algorithm Truncated Energy \cite{Abbasi:2012wht} and the true energy loss of the muon (left), the muon energy when entering the detector (middle) and the primary neutrino energy (right) obtained from Monte Carlo. The spectral shape assumed in these plots is the best-fit superposition of atmospheric and astrophysical neutrino fluxes from this analysis. Each column has been normalized individually to $1$ for a better visibility of the reconstruction uncertainties.}
%\Comment{(Comment)}{we are corrently preparing a version showing TRUE dE/dx vs RECO dE/dx as an alternative.}
%\Comment{AS}{Currently under discussion to show TRUE dE/dx vs RECO dE/dx as a resolution plot. Opinions?}
  \label{img:EnergyResolution}
\end{figure*}

The angular and energy resolution of events in the final event selection are illustrated in Figs.~\ref{img:ZenithResolution} and \ref{img:EnergyResolution}. Ninety percent of the conventional atmospheric neutrinos are reconstructed within $3^{\circ}$ of their true direction and $50\%$ within $1^{\circ}$. As more energetic tracks deposit more light in the optical sensors, the resolution is better for the harder energy spectrum than for atmospheric neutrinos.

A reconstruction of the neutrino energy is challenging because the detector only observes the deposited energy loss for a throughgoing muon \cite{energypaper}. This reconstruction is based on the measurement of the amount of light deposited along the track \cite{Abbasi:2012wht}.
In this algorithm the  40\% of the DOMs with the largest measured  charge 
have been  removed for the energy loss estimation (truncated energy loss).
This leads to an  underestimation of the total energy loss, however
this observable is less sensitive to stochastic fluctuations in the energy loss.
 Figure \ref{img:EnergyResolution} shows that the reconstructed truncated energy loss is well correlated to the true muon energy loss. This reconstructed  energy loss is further correlated to the total muon energy, which is further correlated to the initital neutrino energy. The uncertainty of this relation increases with energy due to the stochastic nature of energy loss processes, and due to the fact that neutrinos may produce high-energy muons far from the detector which will be recorded with lower energy after travelling through the rock and ice.

%Based on the amount of light deposited along the track, the energy loss per unit path length of the muon is estimated \cite{Abbasi:2012wht}. However, this deposited energy loss is correlated to the initial neutrino energy, which is illustrated by Fig.~\ref{img:EnergyAndZenithResolution}. The spread increases with energy due to the stochastic nature of energy loss processes, and due to the fact that neutrinos may produce high-energy muons far from the detector which will be recorded with lower energy after travelling through the rock and ice.

\begin{figure}
  \includegraphics[width=\linewidth]{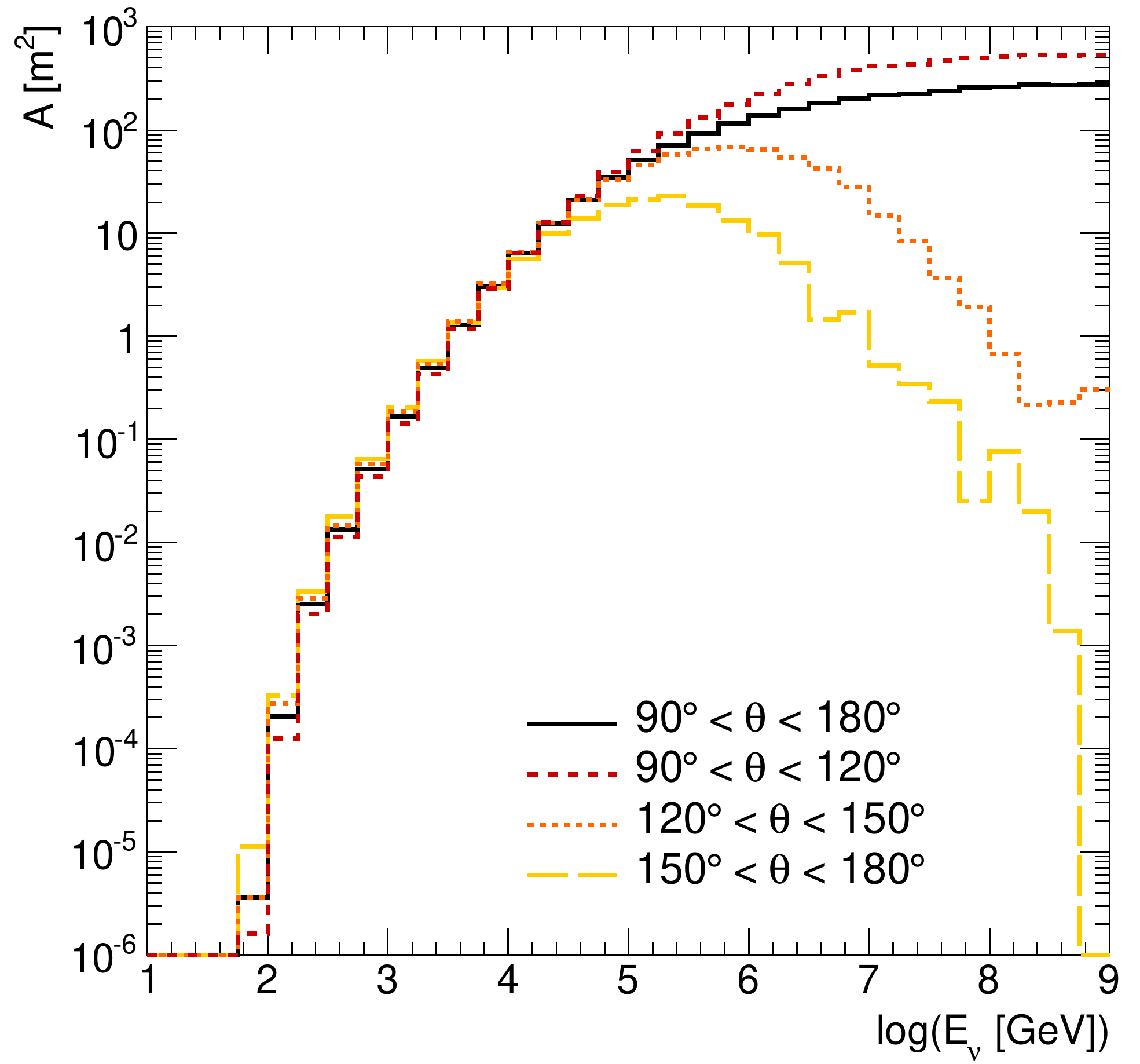}
  \caption{Neutrino effective area averaged for $\nu_{\mu}$ and $\overline \nu_{\mu}$ of the final event selection for different zenith bands.}
  \label{img:EffectiveArea}%
\end{figure}

%The sensitivity of the analysis can be expressed as a neutrino effective area $A$, which is a function of the neutrino energy and direction. 
The detection efficiency of a data sample can be expressed in terms of an effective area $A_{\textrm{\tiny{eff}}}$. For a neutrino flux arriving at the Earth's surface $\Phi (E_{\nu})$, the mean rate $R$ of neutrino events within a solid angle $\Omega$ and an energy interval $\Delta E_{\nu}$ is proportional to this area:
\begin{equation}
R = \int d\Omega \int_{\Delta E_{\nu}} dE_{\nu} A_{\textrm{\tiny{eff}}}(E_{\nu}, \theta) \Phi (E_{\nu}, \theta).
\end{equation}
The effective area for this data sample is shown in Fig.~\ref{img:EffectiveArea}. Overall, the effective area increases with energy. However, at very high energies, neutrinos are absorbed inside the Earth. This reduces the effective area in particular for vertically upward-going events. High-energy neutrinos are therefore expected to arrive predominantly from horizontal directions. The energy threshold of this analysis is around $100\,$GeV.

\section{Analysis method}
\label{likelihood}

\begin{figure*}
  \includegraphics[width=0.32\textwidth]{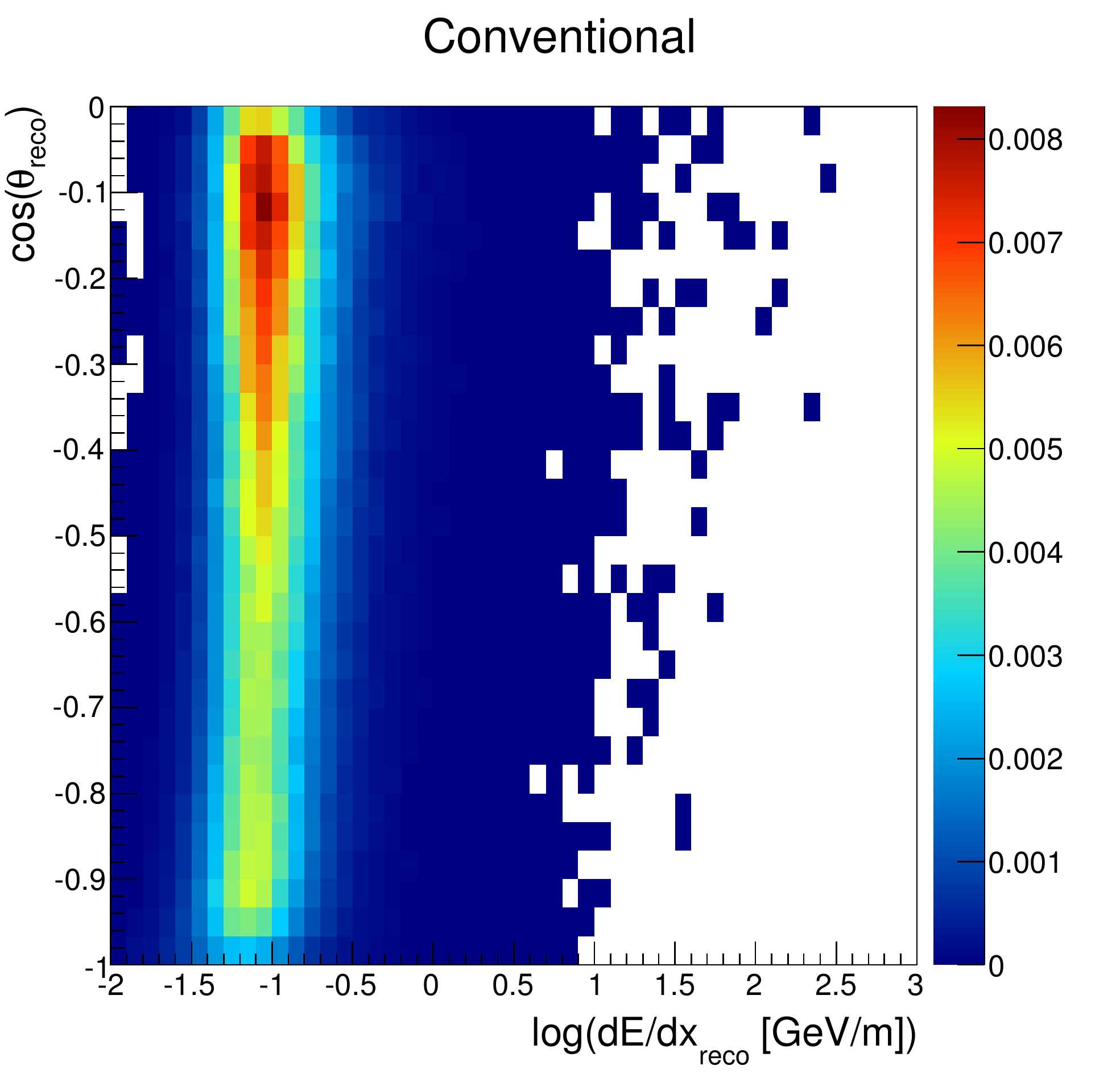}
  \includegraphics[width=0.32\textwidth]{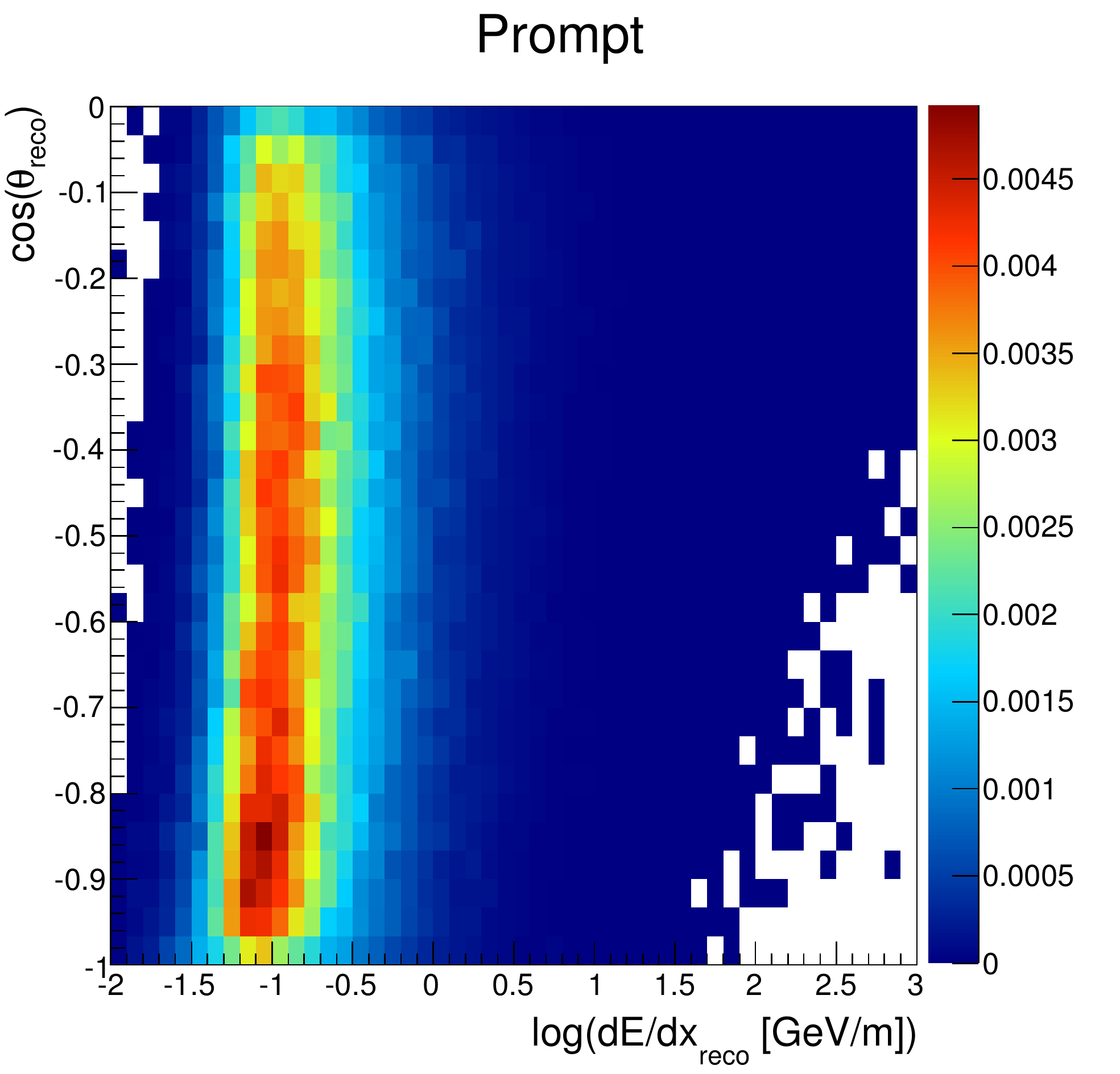}
  \includegraphics[width=0.32\textwidth]{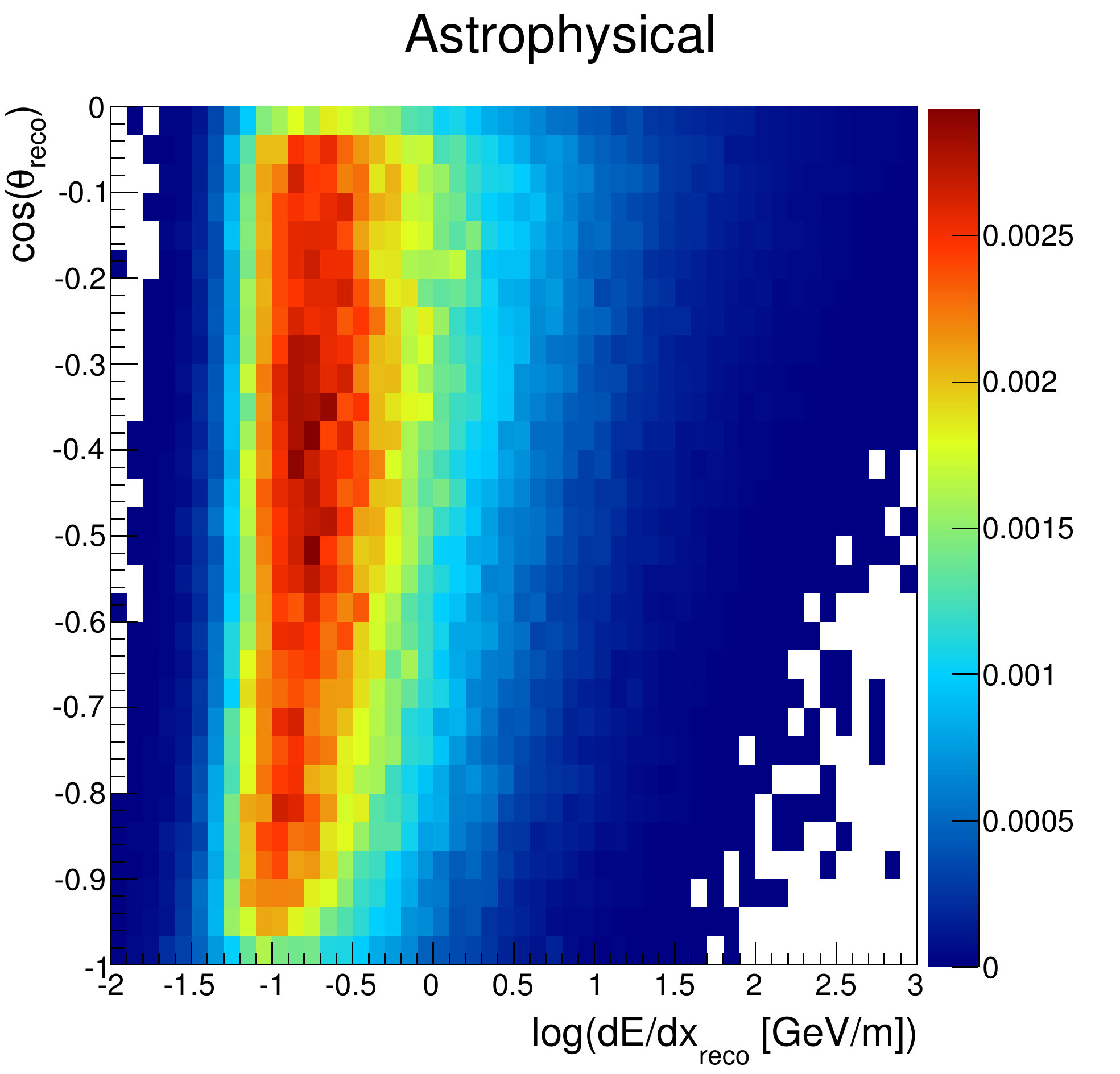}
  \caption{Probability density functions in reconstructed truncated energy loss and cosine zenith angle for conventional atmospheric, prompt atmospheric and astrophysical neutrinos at neutrino level, each normalized to sum over all bins to unity.}%
  \label{img:pdfs}
\end{figure*}

\begin{figure}
  \includegraphics[width=\linewidth]{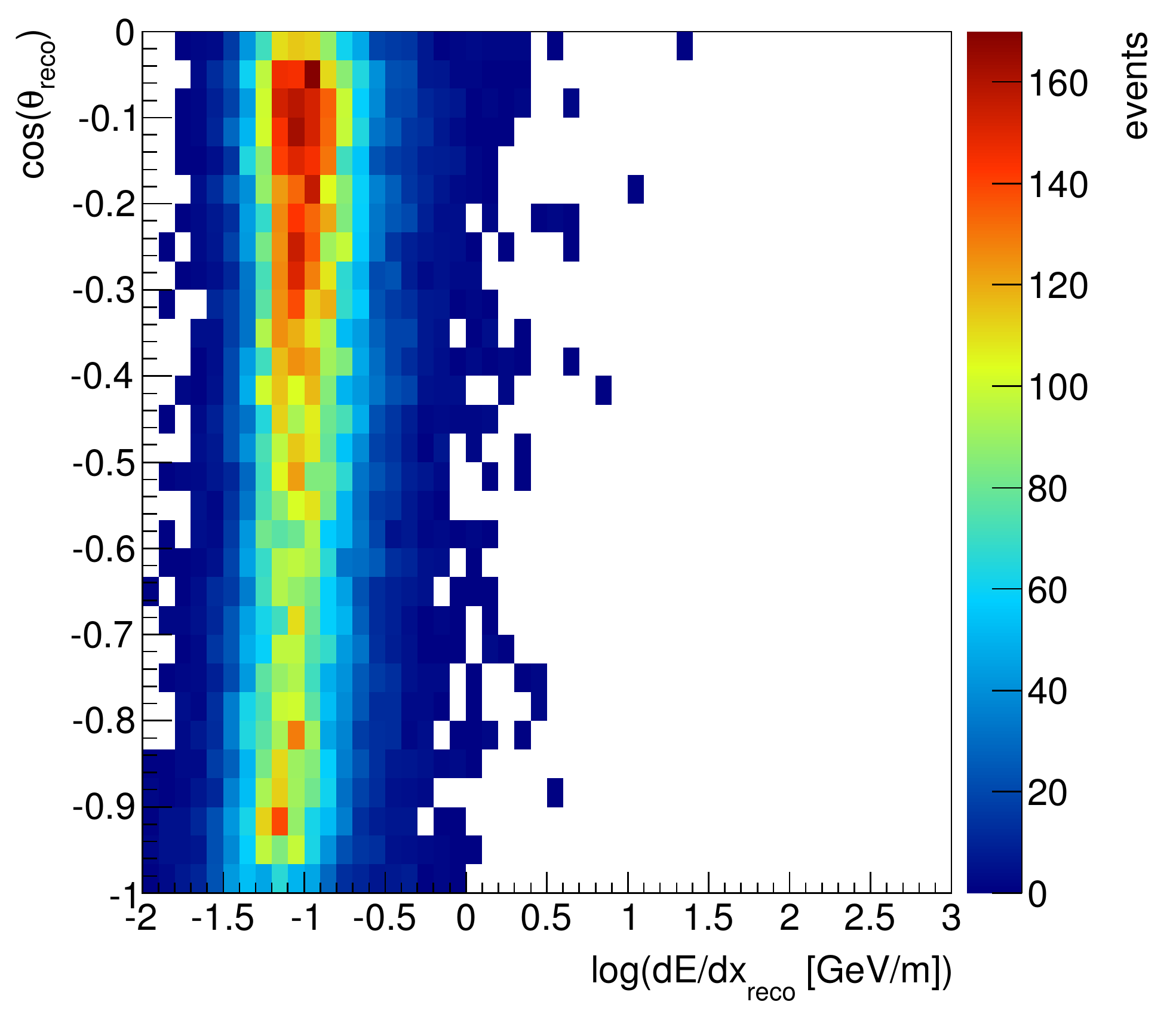}
  \caption{Reconstructed 2-di\-men\-sio\-nal distribution of truncated energy loss and zenith angle for the final event selection with 21943 events.}%
  \label{img:Data2D}%
\end{figure}

The expected distributions of deposited energy and zenith direction as well as their correlation differ for astrophysical, prompt and conventional neutrino signals. The two-dimensional probability density functions, derived from simulation, are displayed in Fig.~\ref{img:pdfs}. In order to identify potential signal components among the background of conventional atmospheric neutrinos, the final neutrino data sample, which is shown in Fig.~\ref{img:Data2D}, is analyzed with a global likelihood fit, determining a best-fit contribution of each component that is statistically consistent with the observed experimental data.

The likelihood $\mathcal{L}$ is the product of likelihoods $\mathcal{L}_{ij}$ for all bins $i, j$ in energy and zenith angle. The likelihood formulation chosen here is a conditional likelihood taking into account that both experimental and simulation data consist of finite statistics \cite{Chirkin:2013lya}. Here, the summed content in each bin $i, j$ consists of $d_{ij}$ experimentally observed 
data counts, and $s_{ij}$ simulated counts, obtained from  simulation with livetime different by a factor $n_s$  to 
 the actual experiment.
   Then, the likelihood is defined by the ratio of the conditional binomial probabilities that the observed sum of simulation and data $s_{ij} + d_{ij}$ for each bin originates from different per-bin expectations $\mu_{s, ij} = s_{ij}/n_s$ and $\mu_{d, ij} = d_{ij}$, and the probability that they originate from the same true values $\mu_{ij} = \mu_{s, ij} = \mu_{d, ij}$. This likelihood $\mathcal{L}_{ij}$ is derived in Ref.~\cite{Chirkin:2013lya} to be
\begin{equation}
\mathcal{L}_{ij} = \left( \frac{\mu_{ij}}{s_{ij}/n_{s}} \right)^{s_{ij}} \cdot \left( \frac{\mu_{ij}}{d_{ij}} \right)^{d_{ij}}.
\label{eq:dimalikelihoodperbin}
\end{equation}
Maximizing this likelihood results in the per-bin expectations $\mu_{ij}$ which agree best with simulation and experimental data. In case of weighted simulation, Eq.~\ref{eq:dimalikelihoodperbin} has to be modified according to Ref.~\cite{Chirkin:2013lya}. It is shown in
\cite{Chirkin:2013lya} that for infinite simulation statistics, this likelihood converges to a  saturated Poissonian likelihood ratio statistic $\mathcal{L}_{ij} = (s_{ij}/n_s)^{d_{ij}} \exp(-s_{ij}/n_s) /( (d_{ij})^{d_{ij}}\exp(-d_{ij}) )$, testing the observed data counts as originating from the exact simulation predictions. However, even with 
a much larger simulation statistics than experimental data, the assumption of infinite simulation statistics is not valid for certain regions of the two-dimensional plane (see Fig.~\ref{img:pdfs}). Hence, the inclusion of the finite simulation statistics into the likelihood formulation has been found to improve the sensitivity of this analysis by about $10$\,\%.

The per-event expectation $\mu_{s, ij}$ for simulation is the sum of the astrophysical, prompt and conventional atmospheric neutrinos to bin $(i, j)$\begin{equation}
\mu_{s, ij} = N_c \cdot p_{ij, c} + N_p \cdot p_{ij, p} + N_{a} \cdot p_{ij, a},
\label{eq:mu}
\end{equation}
where the factors $p_{ij}$ are defined from the probability density functions (see Fig.~\ref{img:pdfs}). The normalization contants $N_c$, $N_p$ and $N_a$ are the parameters describing signal and background contributions to the data sample which are derived from the fit. 
%For the limit of infinite simulation statistics, the best description is $\mu_{ij} = \mu_{s, ij}$ and the likelihood turns into the standard Poissonian formulation \cite{Chirkin:2013lya}.

This formulation is designed to mitigate the effects of finite simulation statistics, which can appear with the use of two-dimensional histograms.
Here, this is particularly evident in the case of the atmospheric neutrino background simulation, (leftmost plot, Fig.~\ref{img:pdfs}).
% for the higher energies, because of the steeply falling energy distribution.
This simulation is based on a reweighting of an $E^{-2}$ source spectrum, and provides good statistics at lower energies, compared to a choice of an $E^{-1}$ based simulation, which, while providing a better estimate of the background at high energies, was found to provide insufficient statistics to describe the zenith angle distribution of conventional atmospheric neutrinos at lower energies. In future analyses, a weighted combination of
simulation sets will be employed to exploit the best features of each, in particular, to improve the background estimate in the signal region.

Systematic uncertainties play an important role in this analysis (see section~\ref{systematics}) and are included as nuisance parameters. These are additional fit parameters which prevent a bias of the signal fit result due to systematic effects. These parameters are penalized by Gaussian prior probabilities, which reflect the range of uncertainty, centered around the expectation value. Some systematic uncertainties, \mbox{e.g.}\,the assumed cosmic-ray parameterization, cannot be easily parameterized as continuous free fit parameters. In such cases, the corresponding uncertainty is still taken into account, as a discrete nuisance parameter. Then, the fit is repeated with 
%different probability density functions for different global settings of 
all discrete settings of the respective systematic uncertainty and the global likelihood maximum is chosen as the best fit. This implementation of nuisance parameters allows the data to constrain these combined effects while simultaneously fitting for possible signals. However, concurrent with the goal of achieving a highly unbiased result for the physics parameters, we note that these nuisance parameters can be highly correlated in their effect on the fitted observables and thus their resulting fit values cannot necessarily be interpreted individually as a measured physical value.

\section{Systematic uncertainties}
\label{systematics}

The determination of a possible signal component through the likelihood fit is based on the comparison of experimental data to simulation and therefore relies on a precise and well understood simulation of signal and background neutrinos. Systematic uncertainties, which affect the efficiency of the selection in the two-dimensional distribution of reconstructed energy loss and arrival direction, are therefore critical.
As a prerequisite, all quality criteria, described in section \ref{eventselection}, have been studied individually to check their agreement between simulation and experiment and their robustness against known systematic uncertainties.

The relevant systematic uncertainties can be grouped into two different categories: the first category (I) is neutrino detection uncertainties and includes uncertainties in the simulation of neutrinos, their interaction and production of secondary particles, propagation in the detection volume, and detector response. The second group (II) consists of uncertainties in the theoretical prediction of energy and zenith angle distributions of atmospheric background neutrinos, such as the normalization, spectral index and knee of the cosmic-ray spectrum and the pion-kaon ratio in air showers.\\ 

\subsection*{{I. Neutrino detection uncertainties}}

\begin{figure}
  \includegraphics[width=\linewidth]{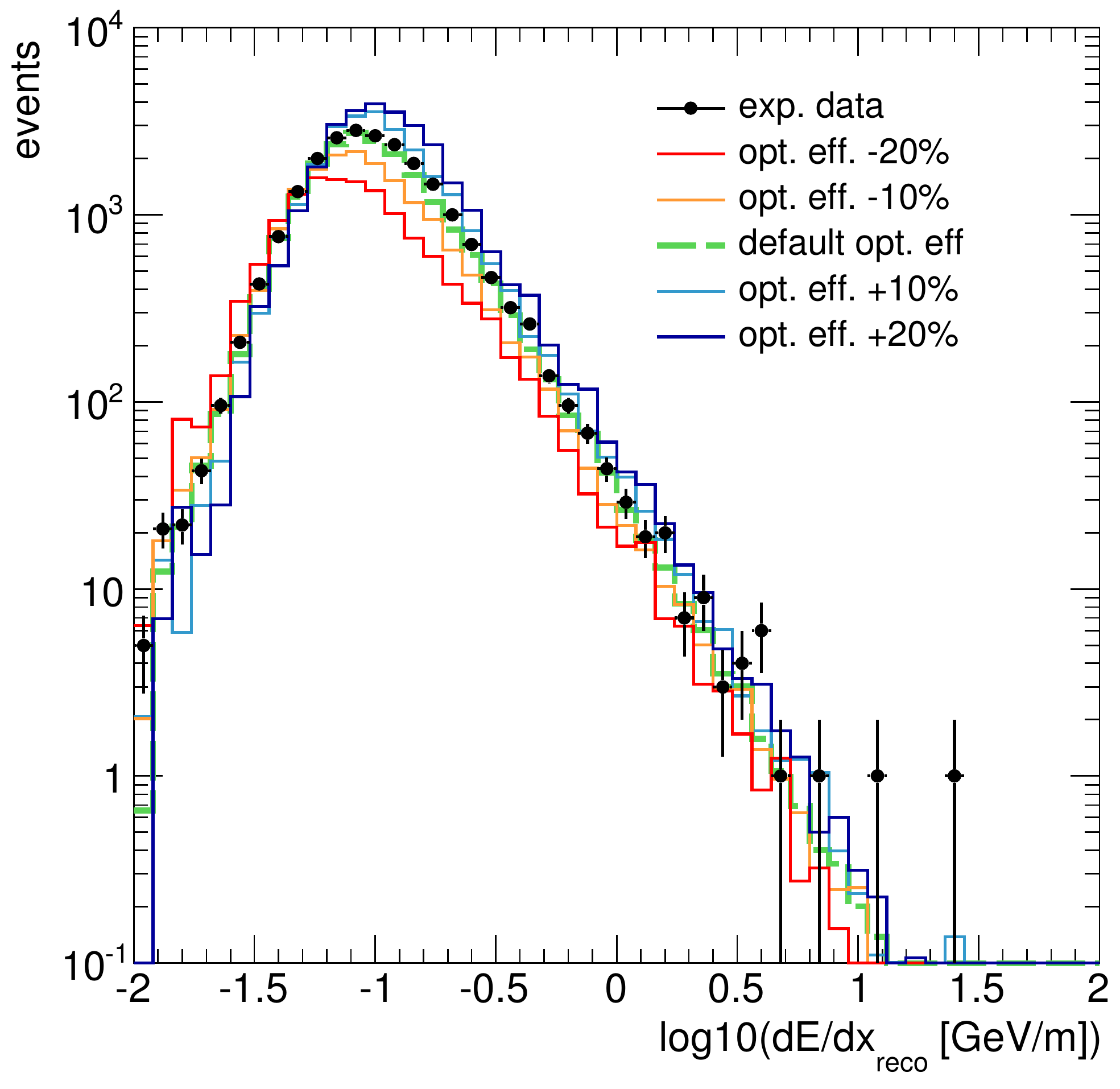}
  \caption{The truncated energy loss distribution of events in the final event selection in comparison to simulations of conventional atmospheric neutrinos (HKKMS07 + H3a) with different settings for the optical efficiency.}%
  \label{img:opteff}%
\end{figure}

\begin{itemize}
\item{Optical efficiency of the detector\\
The optical efficiency includes all uncertainties concerning light production and light detection in the detector. These are the number of produced Cherenkov photons for each propagating charged particle, in particular muons, the overall optical transparency of the ice, the ice properties inside the re-frozen holes around IceCube strings, the photon detection efficiency of the photomultipliers, the photon detection efficiency of the total optical module including glass and gel transparency, i.e.~its effective aperture, and the shadowing of photons by detector components, i.e.~cables and the mu-metal grids. All these factors influence how bright a simulated neutrino appears in the detector. The brightness of an event is the basic information for every energy reconstruction and therefore the uncertainty of the optical efficiency results in an uncertainty on the reconstructed energy scale. Additionally, it affects the normalization and slope of the energy loss distribution, as shown in Fig.~\ref{img:opteff}. The effect has been parameterized and is implemented as a continuous nuisance parameter assuming a Gaussian uncertainty of $15\%$. %\cite{DOMEffUncertaintyPaper}. 
%The influence is correlated to the influence of flux normalization and spectral index.
}

\item{Neutrino-nucleon cross sections\\
The influence of uncertainties in the differential neutrino-nucleon cross sections on the observables of this analysis is estimated through a comparison of neutrino simulations with different cross sections models \cite{Lai:1999wy,CooperSarkar:2007cv,CooperSarkar:2011pa}. While the differences between these models are energy dependent, the effect on the observables is marginal and correlated to other uncertainties. It is therefore neglected here (see Tab.~\ref{tab:uncertainties}).
}

\item{Uncertainties on muon energy loss processes\\
The dominant energy loss processes in the energy range of this analysis are bremsstrahlung, pair production, and photo-nuclear interactions. Theoretical uncertainties are $2\%$ for bremsstrahlung and $2.3\%$ for pair production \cite{Chirkin:2004hz}. The uncertainties for photo-nuclear interactions are of the order of $5\%$ \cite{Sokalski:2002dk, Chirkin:2004hz}, but photo-nuclear interactions contribute less to the total energy energy loss of the muon than bremsstrahlung. Neutrino simulations with varied cross sections of the order of the given uncertainties showed no significant effect on the observables. Expected effects are correlated to other uncertainties and the uncertainty is therefore neglected here (see Tab.~\ref{tab:uncertainties}).
}

%\item{Neutrino-nucleon cross sections and cross sections for muon energy loss processes\\
%The influence of uncertainties in the differential neutrino-nucleon cross sections on the observables of this analysis is estimated through a comparison of neutrino simulations with different assumptions of cross sections models \cite{Lai:1999wy,CooperSarkar:2007cv,CooperSarkar:2011pa}. Theoretical uncertainties for the dominant energy loss processes in the energy range of this analysis are $2\%$ for bremsstrahlung and $2.3\%$ for pair production \cite{Chirkin:2004hz}. The effect on the observables is marginal and affects the normalization and slope of the energy loss distribution. 
%These effects are correlated to the uncertainties in flux normalization, spectral index and optical efficiency.
%}

\item{Optical properties of Antarctic ice\\
Photons produced by secondary particles in the detection volume are subject to scattering and absorption during propagation to the DOMs. The optical properties of the Antarctic ice have been estimated using calibration light sources inside the ice following two approaches \cite{ackermann2006optical, Aartsen:2013rt} and show a spatial dependence in particular in the vertical direction. The influence on the observables due to both ice models (SPICE Mie, WHAM!) is not parameterizable in terms of fundamental parameters and the ice model is therefore taken into account as a discrete nuisance parameter. 
The overall influence of the ice model uncertainty is found to be largely correlated to the flux normalization and the spectral index.
}
\end{itemize}

\begin{table*}[t]\index{typefaces!sizes}
  \footnotesize%
  \caption{Summary of systematic uncertainties and their implementation into this analysis. For continous parameters constraints are implemented as a Gaussian prior around the default value with a standard deviation $\sigma $.}
  \begin{center}
    \begin{tabular}{l|ccccc}
      \hline
        uncertainty & effect on observables & correlated with & implementation & default value &  constraint\\
      \hline
      \hline
	\multirow{2}{*}{optical efficiency} & shape of energy dist. & flux normalization & \multirow{2}{*}{continuous} & \multirow{2}{*}{$1.0$} & \multirow{2}{*}{$\sigma =\pm 0.15$}\\
	 & norm. of energy and zenith dist. & spectral index & & & \\
      \hline
	\multirow{2}{*}{neutrino-nucleon} & \multirow{2}{*}{marginal effect on slope and norm. of} & optical efficiency & \multirow{3}{*}{neglected} & & \\
	\multirow{2}{*}{cross sections} & \multirow{2}{*}{energy and zenith distributions} & spectral index & & & \\
	 & & flux normalization & & & \\
      \hline
	\multirow{2}{*}{muon energy loss} & \multirow{2}{*}{marginal effect on slope and norm. of} & optical efficiency & \multirow{3}{*}{neglected} & & \\
	\multirow{2}{*}{cross sections} & \multirow{2}{*}{energy and zenith distributions} & spectral index & & & \\
	 & & flux normalization & & & \\
      \hline
	\multirow{2}{*}{ice model} & slope and norm. of energy & flux normalization & \multirow{2}{*}{discrete} & & SPICE Mie/\\
	 & and zenith distribution & spectral index & & & WHAM!\\
      \hline
      \hline
	\multirow{2}{*}{flux normalization} & norm. of energy and & optical efficiency & \multirow{2}{*}{continuous} & \multirow{2}{*}{$1.0$} & \multirow{2}{*}{$\sigma = \pm 0.3$}\\
	 & zenith angle distributions & spectral index (weakly) & & & \\
      \hline
	knee of the & slope of energy dist. & \multirow{2}{*}{spectral index} & \multirow{2}{*}{discrete} & & H3a/\\
	cosmic-ray spectrum & minor effect on zenith angle & & & & poly-gonato\\
      \hline
	change in cosmic-ray & slope of energy dist. & optical efficiency & \multirow{2}{*}{continuous} & \multirow{2}{*}{$\Delta \gamma = 0$} & \multirow{2}{*}{$\sigma =\pm 0.1$}\\
	spectral index & minor effect on zenith angle & flux normalization (weakly) & & & \\
      \hline
	\multirow{2}{*}{pion-kaon ratio} & slopes of energy and & & \multirow{2}{*}{continuous} & \multirow{2}{*}{$1.0$} & \multirow{2}{*}{$\sigma =\pm 0.1$}\\
	 & zenith angle distributions & & & & \\
      \hline
    \end{tabular}
  \end{center}
  \label{tab:uncertainties}
\end{table*}

\subsection*{\raggedright{II. Atmospheric neutrino flux uncertainties}}

\begin{itemize}
\item{Flux normalization\\
The uncertainty on the normalization of the conventional atmospheric neutrino background is assumed to be about $30\%$ \cite{Honda:2006qj}. It is implemented as a continuous nuisance parameter with a Gaussian constraint on the conventional flux of atmospheric neutrinos. 
%The flux normalization is correlated to all other uncertainties, which affect the overall normalization, e.g.~the optical efficiency. 
The prompt neutrino flux normalization is not constrained, because it is treated as a signal parameter.
}

\item{Knee of the cosmic-ray spectrum\\
The characteristic shape of the all-particle cosmic-ray energy spectrum with a break at the knee with an energy of $3\,$PeV is caused by the superposition of the spectra of different nuclei. The neutrino fluxes based on the cosmic-ray parameterizations H3a and poly-gonato (see Appendix) change the atmospheric background expectation in a non-trivial way (see Fig.~\ref{img:neutrinofluxeswithknee} in the Appendix): the neutrino flux expectation increases towards higher energies but is strongly reduced at highest energies due to the cosmic-ray knee. The different cosmic-ray spectra are implemented into the fit as a discrete nuisance parameter.

}

\item{Cosmic-ray spectral index\\
Since conventional and prompt atmospheric neutrinos are produced by cosmic rays hitting the atmosphere, their energy spectrum directly depends on the energy spectrum of cosmic rays. As discussed above, the cosmic-ray nucleon spectrum, which is not a simple power law, is relevant for the neutrino flux estimation. The overall uncertainty on the spectral index is implemented as a continuous uncertainty shifting the total spectrum by $\Delta \gamma$ relative to the parameterizations of the cosmic-ray composition models discussed above. The constraint of $4\%$ is estimated based on differences between established cosmic-ray flux parameterizations \cite{Gaisser:2012zz,hoerandel2003knee}. The sign of $\Delta \gamma $ is defined such that a positive value corresponds to a softer spectrum. 
%The uncertainty is correlated to other uncertainties which affect the energy scale such as the optical efficiency, and is weakly correlated to the flux normalization.
}

\item{Pion-kaon ratio\\
The relative pion to kaon contribution to neutrino production in air showers is the main uncertainty affecting the zenith angle distribution. It is defined here as the ratio of the integrated pion and kaon neutrino flux contribution to the total neutrino flux in this data sample from Eq.~\ref{eq:gaisserneutrinoflux}. In the analysis, it is implemented as a continuous nuisance parameter with a Gaussian constraint of $10\%$. This corresponds to a $3\%$ uncertainty in the vertical to horizontal flux ratio, which is estimated from theoretical calculations \cite{Honda:2006qj}.
}

\end{itemize}

Many of these uncertainties are highly correlated in their effect on the analysis observables (see Tab.~\ref{tab:uncertainties} and Fig.~\ref{img:correlations}). The purpose of the implementation of these uncertainties as nuisance parameters in the fit is to avoid the mis-interpretation of deviations between experiment and simulation due to these uncertainties as an astrophysical or prompt neutrino signal. This effect has been checked for by using simulation-based data challenges prior to the final analysis. Ensembles of experiments were generated with varying nuisance parameter settings, and then the entire fitting procedure was applied with varying assumptions on the ranges of the nuisance parameters in the fit. It was found that the procedure is very robust against nuisance parameter assumptions, and that the chance of mis-interpreting systematic deviations as a signal is low.

As discussed above, the correlation between different nuisance parameters does not permit a precise determination of the corresponding physics parameters from the fit. A nuisance parameter can be absorbed by another free floating nuisance parameter describing different uncertainties
if the influence on the observables is correlated. Examples are: the uncertainty of quantum efficiency of optical sensors, which is correlated to the Cherenkov light yield uncertainty; the effects of the uncertainties in the cross sections for neutrino-nucleon interactions and muon energy loss, which are fully absorbed by other parameters.
%(CW) this is I think a better example
%uncertainties 
%which is large and for.\\
For such cases only a single parameter with the combined uncertainty has been implemented into the analysis to ensure good numerical stability of the fit.

A summary of all systematic uncertainties and their implementations is given in Tab.~\ref{tab:uncertainties}.

\section{Results}
\label{results}

\subsection{Likelihood fit results}

\begin{figure*}
  \begin{center}
  \includegraphics[width=0.48\textwidth]{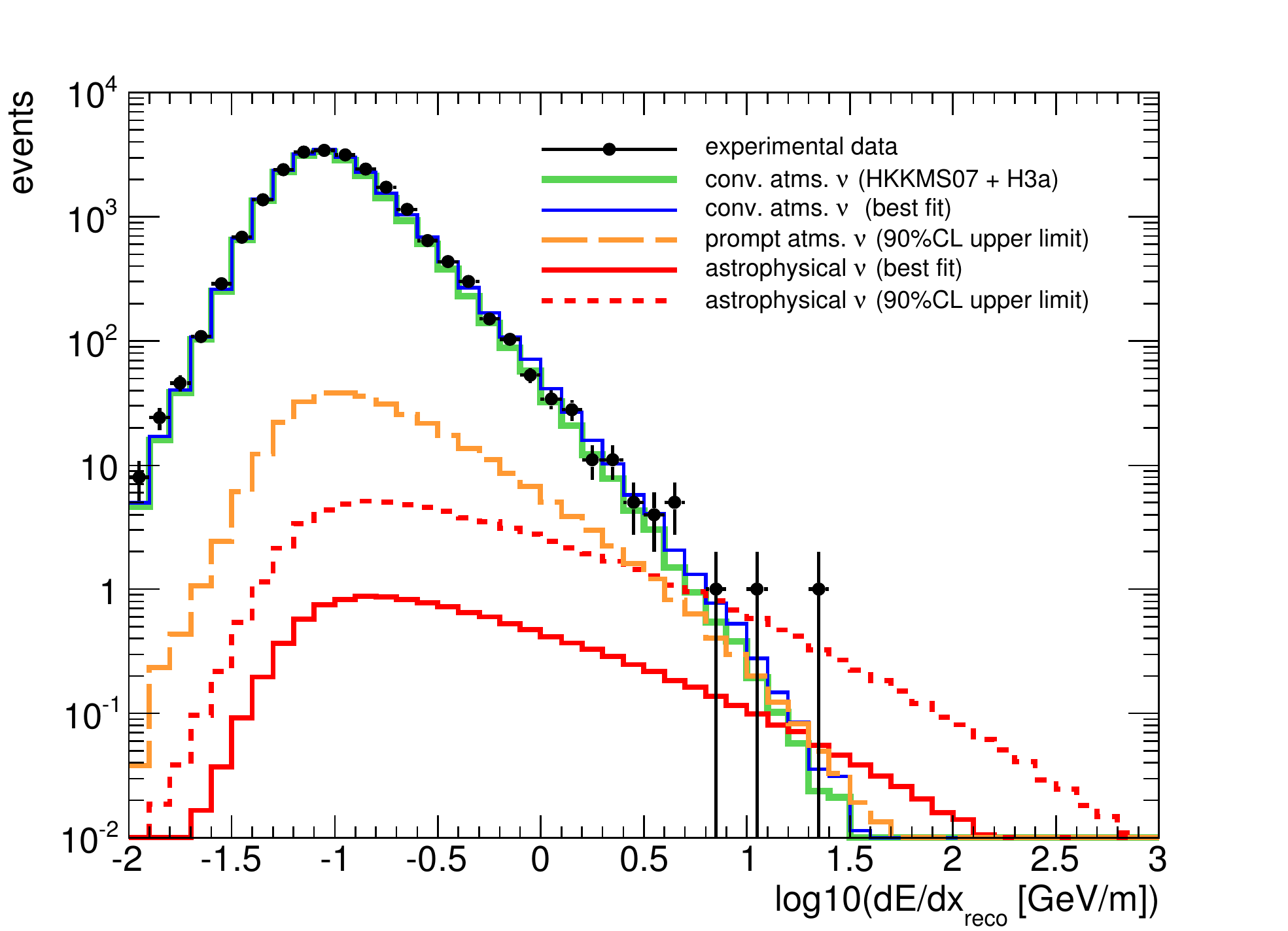}
  \includegraphics[width=0.48\textwidth]{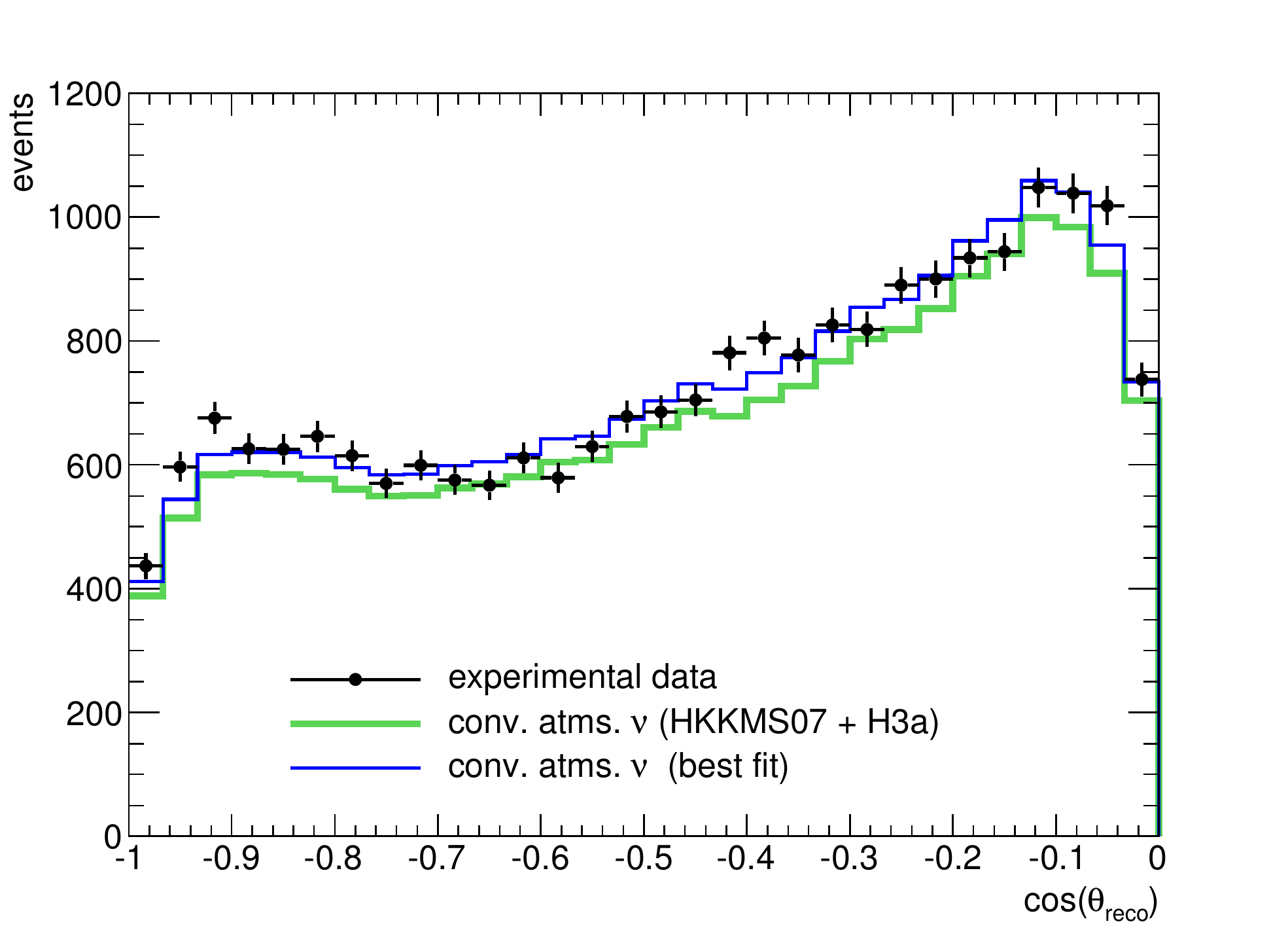}
  \end{center}
%\Comment{CW}{We could also show a version without the default MC. Plots are available. Opinions?}
%envelope\_justlines\_nomc\_energy and envelope\_justlines\_nomc\_zenith)
  \caption{Truncated energy loss and zenith angle distribution of the final neutrino data sample in comparison to the simulation of conventional atmospheric neutrinos with default nuisance parameters (green thin line) and the best fit conventional atmospheric neutrino (blue thick line), the best fit (red solid line) and upper limit astrophysical spectra (red dashed line) and the upper limit prompt neutrino spectrum (orange long-dashed line).}
  \label{img:EnergyAndZenith}
%\Comment{CW}{Zenith figure looks inconsistent to Anne's thesis. See my comment to Shigeru}
\end{figure*}

The two-dimensional distribution of reconstructed truncated energy loss and zenith angle for the high-purity experimental neutrino sample %, containing 21943 events
is shown in Fig.~\ref{img:Data2D}. The best fit of this distribution as a superposition of the three neutrino components, astrophysical, prompt atmospheric and conventional atmospheric (see section~\ref{likelihood}), with the nuisance parameters allowed to float within constraints (see Tab.~\ref{tab:uncertainties}), is summarized in Tab.~\ref{tab:FitResults}. The best fit for the astrophysical component is a flux of 
\begin{equation}
E_{\nu}^2 \Phi (E_{\nu}) = 0.25\cdot 10^{-8}\,\textrm{GeV}\,\textrm{cm}^{-2}\,\textrm{s}^{-1}\,\textrm{sr}^{-1},
\end{equation}
and the best fit of the prompt atmospheric flux is zero. The projected distributions of the truncated energy loss and zenith angle are shown in Fig.~\ref{img:EnergyAndZenith}.

\begin{figure}
  \includegraphics[width=0.7\linewidth]{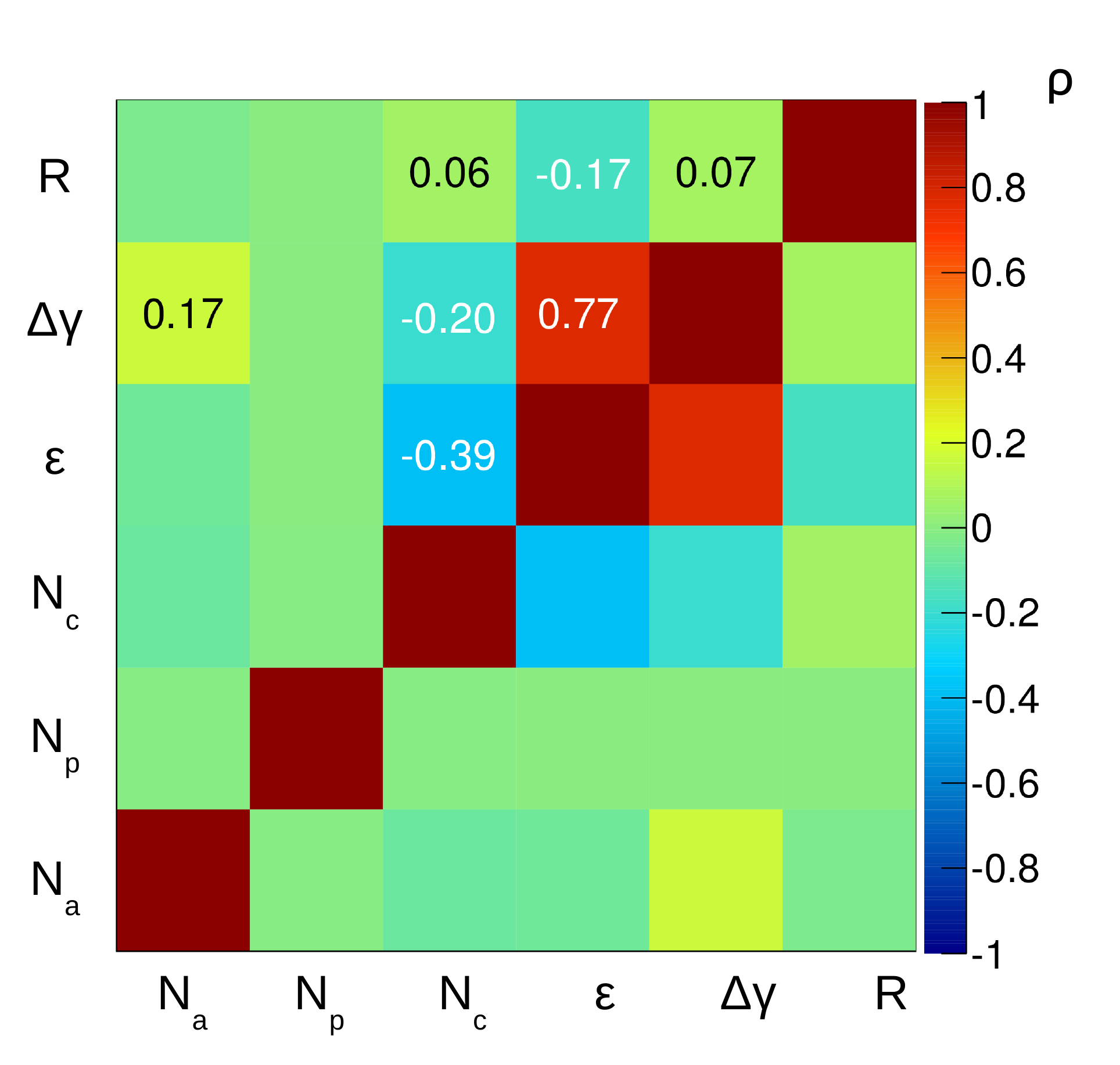}
  \caption{Correlation coefficients between continuous fit parameters in the fit of the experimental data. The parameters $N_a$, $N_p$, and $N_c$ are the normalizations of astrophysical, prompt and conventional atmospheric neutrinos (see Eq.~\ref{eq:mu}). The other parameters are the continuous nuisance parameters of the fit, i.e.~the optical efficiency $\epsilon$, the change in spectral index $\Delta \gamma$ and the pion-kaon ratio scaling factor $R$ (see section \ref{systematics} and Tab.~\ref{tab:uncertainties}).}%
  \label{img:correlations}%
\end{figure}

All nuisance parameter best fit values are consistent with expectations. The correlations between the continuous signal and nuisance fit parameters are shown in Fig.~\ref{img:correlations}.
% redundant Due to a high correlation between nusiance parameters, the resulting fit values cannot be directly interpreted as an atmospheric neutrino flux measurement. As an example, the energy loss distribution is shown in Fig.~\ref{img:opteff} for the experimental data and different values of the optical efficiency. 
Table~\ref{tab:FitResults} shows statistical errors on the parameters of the fit, which represent the ranges allowed by the fit. 
The statistical error on the signal fit comes from a full ensemble construction using likelihood ratios. The statistical errors on the systematic parameters come from a standard $\chi^2$ interpretation of the changes in likelihood from the minimum. The systematic pulls are also shown, which indicate how far the nuisance parameters have moved from their assumed baseline values. The pion-kaon ratio shows the largest pull, a $13\%$ increase in the kaon contribution from the baseline assumption. This increase is not statistically significant and due to the different sensitive energy regions of the respective analyses, cannot be directly compared to the studies of the pion-kaon ratio with atmospheric muons \cite{Grashorn:2009ey}.\\

\begin{table*}\index{typefaces!sizes}
  \footnotesize%
  \caption{Fit results for the fit parameters from the likelihood analysis. The results for the discrete nuisance parameters ``model of optical ice properties'' and ``knee of the cosmic-ray spectrum'' are those models which return the best likelihood value during the fit.}
  \begin{center}
    \begin{tabular}{l|c|c|c}
      \hline
        fit parameter & fit value & stat. error on best fit & systematic pull\\
      \hline
      \hline
        Astrophys. flux [$10^{-8}\,$GeV$\,$cm$^{-2}\,$s${-1}\,$sr$^{-1}$] & $0.25$ & $+ 0.70 - 0.20$ & \\
	Prompt flux $N_p$ [ERS08 + H3a] & $0$ & $ + 2.41$ & \\
      \hline
      \hline
	Optical efficiency $\epsilon$ & $1.00$ & $\pm 0.01$ & $0\sigma$\\
        Model of optical ice properties & SPICE Mie & & \\
      \hline
      \hline
        Conventional flux $N_c$ [HKKMS07 + H3a] & $1.05$ & $\pm 0.02$ & $+0.2\sigma$\\
        Knee of the cosmic-ray spectrum & H3a & & \\
        Change in spectral index $\Delta \gamma$ & $-0.06$ & $\pm 0.02$ & $-0.6\sigma$\\
        Pion-kaon ratio scaling factor $R$ & $1.13$ & $\pm 0.10$ & $+1.3\sigma$ \\
      \hline
    \end{tabular}
  \end{center}
  \label{tab:FitResults}
\end{table*}

The fitted non-zero astrophysical signal flux is found close to the physical boundary. As the fit is constrained to non-negative signal fluxes, the significance of the likelihood ratio was determined using full ensemble constructions. 
%Fig.~\ref{img:Contour} shows the significance of the result in the signal parameter space of astrophysical and prompt neutrino fluxes. 
For this, we define a test statistic for the evaluation of each point $\mathbf{a}$ in the signal parameter space by the likelihood ratio $R$. $R$ is defined as the ratio of the likelihood of best fit signal $\mathbf{\hat{a}}$ and nuisance parameters $\mathbf{\hat{b}}$ and the likelihood of a fit with signal parameters fixed at $\mathbf{a}$ and best fit nuisance parameters $\mathbf{\hat{\hat{b}}}$:
\begin{equation}
\label{eq:LHR}
R = - 2 \ln \frac{\mathcal{L} \left( \mathbf{a}, \mathbf{\hat{\hat{b}}} \right)}{\mathcal{L} \left( \mathbf{\hat{a}}, \mathbf{{\hat{b}}} \right)}.
\end{equation}
The p-values for each signal hypothesis test can be estimated through a comparison to a distribution of $R$ from $N$ simulated ensembles and
the experimental value $R_{\textrm{\tiny{exp}}} $ by the number $N$ of ensemble tests with a larger $R$ value than the experimental result:
\begin{equation}
\label{eq:pvalue}
\textrm{p-value} = \frac{N \left( R > R_{\textrm{\tiny{exp}}} \right)}{N}.
\end{equation}
The hypothesis of zero signal ($\mathbf{a} = (0, 0)$) results in a p-value of $0.032$. This corresponds to a 1-sided significance of $1.8\,\sigma$, a rejection at a 96.8\% confidence level. Testing a range of values leads to the $68\%$ confidence level allowed range for the astrophysical signal
%of $E_{\nu}^2 d\Phi/dE_{\nu} = (0.04 , 0.94) \cdot 10^{-8}\,\textrm{GeV}\,\text%rm{cm}^{-2}\,\textrm{s}^{-1}\,\textrm{sr}^{-1}$.
of 
\begin{equation}
0.04 \le \frac{E_{\nu}^2 \cdot \Phi(E_{\nu})}{10^{-8} \, \textrm{GeV}\,\textrm{cm}^{-2}\,\textrm{s}^{-1}\,\textrm{sr}^{-1}} \le 0.94 \quad .
\end{equation}

%\begin{figure}
%  \includegraphics[width=\linewidth]{fit_contour-placeholder}
%  \caption{Best fit and contours in the parameter space of astrophysical and prompt flux components.}%
%  \label{img:Contour}%
%\end{figure}

Further, the upper limits on the astrophysical and prompt atmospheric muon neutrino fluxes were calculated using the same ensemble method. The upper limits at $90\%\,$confidence level are
\begin{equation}
\frac{E_{\nu}^2 \cdot \Phi_{\textrm{\tiny{astro}}}(E_{\nu})}{ 10^{-8}\,\textrm{GeV}\,\textrm{cm}^{-2}\,\textrm{s}^{-1}\,\textrm{sr}^{-1}} \le 1.44 \quad .
\end{equation}
in the energy range between $34.5\,$TeV and $36.6\,$PeV and
\begin{equation}
\Phi_{\textrm{\tiny{prompt}}} (E_{\nu}) \le 3.8\cdot \Phi_{\textrm{\tiny{ERS08 + H3a}}} (E_{\nu})
\end{equation}
for the baseline model ERS08 + H3a in the energy range between $2.3\,$TeV and $360\,$TeV. The sensitive energy range of the analysis is defined as the energy range which achieves a 5\% worse sensitivity than the full energy range if signal pdfs are constrained from the high and low-energy side respectively. The best-fit and upper-limit projected distributions of the reconstructed energy loss and zenith angle are illustrated in Fig.~\ref{img:EnergyAndZenith}. An astrophysical neutrino flux at the level of the best fit would yield $12$ signal neutrino events in the final neutrino data sample, and a flux at the level of the upper limit would yield $71$ neutrino events. A flux at the prompt upper limit would correspond to $346$ neutrinos in this data sample, which can be compared to $91$ expected prompt atmospheric neutrinos assuming the ERS08 + H3a model.

\subsection{Limits on an astrophysical $E^{-2}$ power law flux}

\begin{figure*}
  \includegraphics[width=\linewidth]{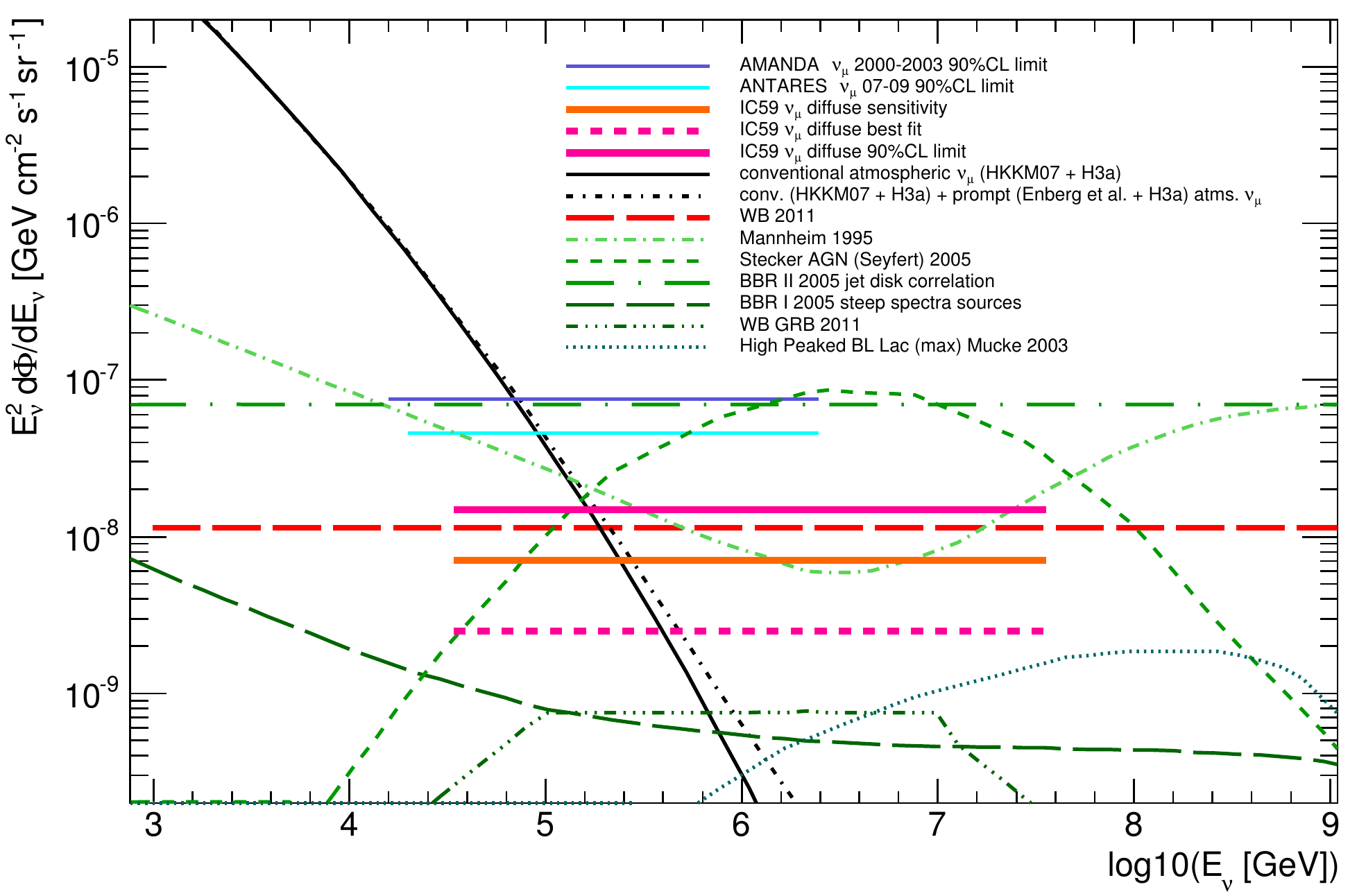}
  \caption{Limit on a $\left( \nu_{\mu} + \overline \nu_{\mu} \right)$ astrophysical $E^{-2}$ flux from this analysis in comparison to theoretical flux predictions and limits from other experiments. The black lines show the expected atmospheric neutrino flux with and without a prompt component (both without the modification of the knee feature). The red dashed line marks the Waxman-Bahcall upper bound \cite{Waxman:1998yy, Waxman:2011hr}. Green dashed lines represent various model predictions for astrophysical neutrino fluxes \cite{Stecker:2005hn, Mannheim:1995mm, Becker:2005ya, Waxman:1997ti, Waxman:1998yy, Muecke:2002bi}. Horizontal lines show limits and sensitivities from different experiments \cite{Achterberg:2007qp, Aguilar:2010ab, Abbasi:2011jx}. The pink solid line is the $90\%$ CL upper limit of this analysis, the orange solid line shows its sensitivity.}%
  \label{img:DiffuseLimit}%
\end{figure*}

Figure \ref{img:DiffuseLimit} compares the upper limit of this analysis with theoretical flux predictions and limits from other experiments. For the first time, this search for astrophysical muon neutrinos with data from the 59-string IceCube detector achieves a sensitivity at a level about 30\% below the Waxman-Bahcall upper bound. However, the upper flux limit of this analysis is about 40\% above the Waxman-Bahcall upper bound. The limit remains a factor of two above the sensitivity of this analysis due to the observed excess of high-energy events, which causes a non-zero astrophysical best-fit flux. Therefore, this optimistic scenario of highly efficient neutrino production in optically thin cosmic-ray sources cannot be excluded.

This result lowers the flux limit of the predecessor experiment AMANDA \cite{Achterberg:2007qp} by a factor of five and is a factor of three below the limit obtained by the ANTARES \cite{Aguilar:2010ab} experiment. This analysis supercedes the result of the IceCube 40-string data analysis \cite{Abbasi:2011jx}. The current study benefited from a number of improvements, in particular, a more efficient neutrino selection, the inclusion of the zenith angle into the fit and a more careful modelling of the primary cosmic-ray spectrum. Furthermore, the previous analysis overestimated its effective area, leading to an underestimated limit. The current 59-string analysis has a substantially higher sensitivity than the 40-string analysis.

%\begin{figure}
%  \includegraphics[width=\linewidth]{%limit_summary_atmosphericwithh3a_noIC40_thickIC59}
%  \caption{Limit on a $\left( \nu_{\mu} + \overline \nu_{\mu} \right)$ astrophysical $E^{-2}$ flux from this analysis in comparison to theoretical flux predictions and limits from other experiments. The black lines show the expected atmospheric neutrino flux with and without a prompt component (both without the modification of the knee feature). The red dashed line marks the Waxman-Bahcall upper bound \cite{Waxman:1998yy, Waxman:2011hr}. Green dashed lines represent various model predictions for astrophysical neutrino fluxes \cite{Stecker:2005hn, Mannheim:1995mm, Becker:2005ya, Waxman:1997ti, Waxman:1998yy, Muecke:2002bi}. Horizontal lines show limits and sensitivities from different experiments \cite{Achterberg:2007qp, Aguilar:2010ab, Abbasi:2011jx}. The pink solid line is the $90\%$ CL upper limit of this analysis, the orange solid line shows its sensitivity.}%
%  \label{img:DiffuseLimit}%
%\end{figure}

IceCube analyses using the channel of cascade-like signatures have recently also found an excess of high-energy neutrino events compared to the number expected from a background of neutrinos of conventional atmospheric origin. These analyses are mostly sensitive to charged-current interactions of electron neutrinos and neutral-current interactions of all neutrino flavors, while the analysis presented here is sensitive mostly to charged-current interactions of muon neutrinos. The energy resolution achieved for contained cascade-like events is better than for a through-going muon track. However, the separation of neutrino signal from atmospheric muon background is more challenging, because of the worse angular resolution achieved in this channel. This in general leads to much smaller event samples than in the track-like channel.

Results from a search for cascade-like high-energy events with the IceCube 40-string detector configuration \cite{eike} showed a high-energy excess of events. The significance of that excess is $2.7\,\sigma$ with respect to the expectation of conventional atmospheric and prompt atmospheric neutrinos. The upper limit derived from that analysis is an all-flavor flux of $E_{\nu}^2 \Phi (E_{\nu}) = 7.46 \cdot 10^{-8}\,\textrm{GeV}\,\textrm{cm}^{-2}\,\textrm{s}^{-1}\,\textrm{sr}^{-1}$ (90\% confidence level). Assuming equal mixing of neutrino flavors when arriving at Earth, that flux is compatible with the best-fit flux and the upper limit derived in this  analysis. 
%%% CW: we do not give numbers of Eike, so we cannot quantify the excess in event numbers
%Based on the same assumption, a flux at the level of the upper limit of this muon neutrino analysis would yield $X$ high-energy neutrinos in the IC40 cascade analysis.

The IceCube collaboration has also reported the observation of 28 high-energy events found in the search for high-energy starting events in the IceCube data taken with the IC79 configuration and the first year of the full 86-string detector \cite{hese}. These 28 events correspond to a $4.1\sigma$ excess with respect to atmospheric background and are interpreted as evidence for an astrophysical all-flavor component of $E_{\nu}^2 \Phi (E_{\nu}) = 3.6 \pm 1.2 \cdot 10^{-8}\,\textrm{GeV}\,\textrm{cm}^{-2}\,\textrm{s}^{-1}\,\textrm{sr}^{-1}$ \cite{hese}. Of these 28 events, only 7 events show a clear track-like signature, the other 21 events have the typical spherical shape of cascade-like events. Assuming again an equal mixing of neutrino flavors, the best-fit flux of the high-energy analysis corresponds to a muon neutrino flux of $E_{\nu}^2 \Phi (E_{\nu}) = 1.2 \cdot 10^{-8}\,\textrm{GeV}\,\textrm{cm}^{-2}\,\textrm{s}^{-1}\,\textrm{sr}^{-1}$ with a cutoff at $2\,$PeV. A fit with an unbroken $E^{-2}$ signal hypotheses as used in the analysis presented here, would yield a slightly lower astrophysical flux normalization. Such a flux level is just below the upper limit set by the muon neutrino search presented here. The best fit astrophysical spectral index, if unconstrained in the fit, is $\gamma = 2.2$. The 28 events predominantly originate from the Southern hemisphere, while the muon neutrino search presented here is only performed for events below the horizon with zenith angles greater than 90 degrees.

\subsection{Limits on astrophysical diffuse neutrino fluxes}

\begin{table*}
  \footnotesize%
  \caption{Model rejection factors and best-fit fluxes in units of the predicted model flux for different theoretical predictions of $\left( \nu_{\mu} + \overline \nu_{\mu} \right)$ astrophysical neutrino fluxes.
%for the following models: neutrinos from AGN cores by Stecker \cite{Stecker:2005hn}, neutrinos from the jets of radio-loud AGN by Mannheim \cite{Mannheim:1995mm}, neutrinos from flat spectrum (BBRII) FR-II radio galaxies and blazars by Becker \textit{et al.}\ \cite{Becker:2005ya}, and high-energy neutrinos from gamma-ray bursts by Waxman and Bahcall \cite{Waxman:1997ti, Waxman:1998yy}.
}
  \begin{center}
    \begin{tabular}{l|l|c|c|c}
      \hline
        \multirow{2}{*}{model} & \multirow{2}{*}{neutrino source} & best fit & \multirow{2}{*}{MRF} & \multirow{2}{*}{energy range}\\
         & & [$\Phi_{\textrm{\tiny{model}}} \left( E_{\nu} \right)$] & & \\
% & likelihood ratio $R$\\
        %Model & neutrino source & best fit [$d\Phi_{\textrm{\tiny{model}}}/dE_{\nu} \left( E_{\nu} \right)$] & MRF \\% & likelihood ratio $R$\\
      \hline
	%Stecker & $0.06$ & $0.33$ & $216\,$TeV to $8.6\,$PeV\\% & 340.988\\
	%Mannheim & $0.13$ & $0.86$ & $28\,$TeV to $2.4\,$PeV\\% & 341.236\\
	%BBRII & $0.03$ & $0.21$ & $73\,$TeV to $8.4\,$PeV\\% & 341.112\\
	%WB GRB & $3.74$ & $21.72$ & $84\,$TeV to $4.3\,$PeV\\% & 341.098\\
	Stecker \cite{Stecker:2005hn} & AGN cores & $0.06$ & $0.33$ & $216\,$TeV to $8.6\,$PeV\\
	Mannheim \cite{Mannheim:1995mm} & jets of radio-loud AGN & $0.13$ & $0.86$ & $28\,$TeV to $2.4\,$PeV \\
	BBRI \cite{Becker:2005ya} & steep spectrum FR-II galaxies and blazars & $3.77$ & $23.07$ & overlap with the atmospheric range \\
	BBRII \cite{Becker:2005ya} & flat spectrum FR-II galaxies and blazars & $0.03$ & $0.21$ & $73\,$TeV to $8.4\,$PeV \\
	Muecke \textit{et al.}~\cite{Muecke:2002bi} & BL Lac objects & $6.83$ & $43.96$ & PeV to EeV energies \\
	WB GRB \cite{Waxman:1997ti, Waxman:1998yy} & gamma-ray bursts & $3.74$ & $21.72$ & $84\,$TeV to $4.3\,$PeV \\
      \hline
    \end{tabular}
  \end{center}
  \label{tab:DiffuseMRFs}
\end{table*}

The experimental data are also compared to various theoretical diffuse neutrino flux models. Best-fit fluxes and upper limits on each of these models are given as a model rejection factor (MRF)\cite{Hill:2002nv}, the ratio between the upper-limit flux assuming the shape of the model prediction and the flux predicted by the model itself. An MRF less than one implies that the model is rejected by the measurement at a confidence level of more than $90\%$. Models with model rejection factors greater than one are constrained by this analysis by less than $90\%\,$CL. In order to calculate the MRFs, the baseline signal hypothesis pdf (see Fig.~\ref{img:pdfs}) has been exchanged by the 2-dimensional energy loss and zenith angle distribution predicted by the corresponding model and the fit has been repeated. The best-fit flux for each model is given in Tab.~\ref{tab:DiffuseMRFs} as the best fit normalization of the model prediction. The MRFs given in Tab.~\ref{tab:DiffuseMRFs} are based on a $\chi^2$-approximation around this best-fit maximum instead of using the computing-intensive confidence interval construction following the Feldman-Cousins approach \cite{PhysRevD.57.3873}.

%Table~\ref{tab:DiffuseMRFs} presents model rejection factors for different diffuse neutrino flux models. 
Three models are excluded by this analysis at $90\%\,$CL: those by Stecker, Mannheim, and the flat spectrum source model by Becker \textit{et al}. The upper limit of this analysis is a factor of $22$ above the Waxman-Bahcall model for GRBs \cite{Waxman:1997ti, Waxman:1998yy}. The analysis presented here cannot constrain this model, but it has already been constrained by dedicated GRB searches with IceCube \cite{Abbasi:2012zw}. There is no sensitivity to the models by Muecke \textit{et al.}~\cite{Muecke:2002bi} and the steep spectrum source model from Becker \textit{et al.}~(BBRI) \cite{Becker:2005ya}. This conclusion is drawn from the fact that the sensitivity to the corresponding model does not worsen by more than 5\% when constraining the signal pdfs in energy. The reason is that the predicted steep spectrum of the BBRI model causes a large degeneracy between the astrophysical signal flux normalization and the atmospheric flux and nuisance parameters. The model by Muecke \textit{et al.}~predicts very low neutino fluxes in the PeV to EeV energy range, which are beyond the reach of current experiments.

%Both predict very low neutrino fluxes, and are not yet in reach of experiments. 

\subsection{Limits on prompt atmospheric neutrino fluxes}

Constraining the prompt atmospheric neutrino flux is challenging, because the prompt flux is hard to distinguish from the shape of the conventional atmospheric flux. In this analysis no indications of a prompt signal are observed.
 The corresponding upper limit on a prompt atmospheric neutrino component 
 can be compared to the results from the AMANDA experiment \cite{Achterberg:2007qp}. Those limits are a factor of seven less constraining than the limits set by the analysis presented here.

\begin{figure}
  \includegraphics[width=\linewidth]{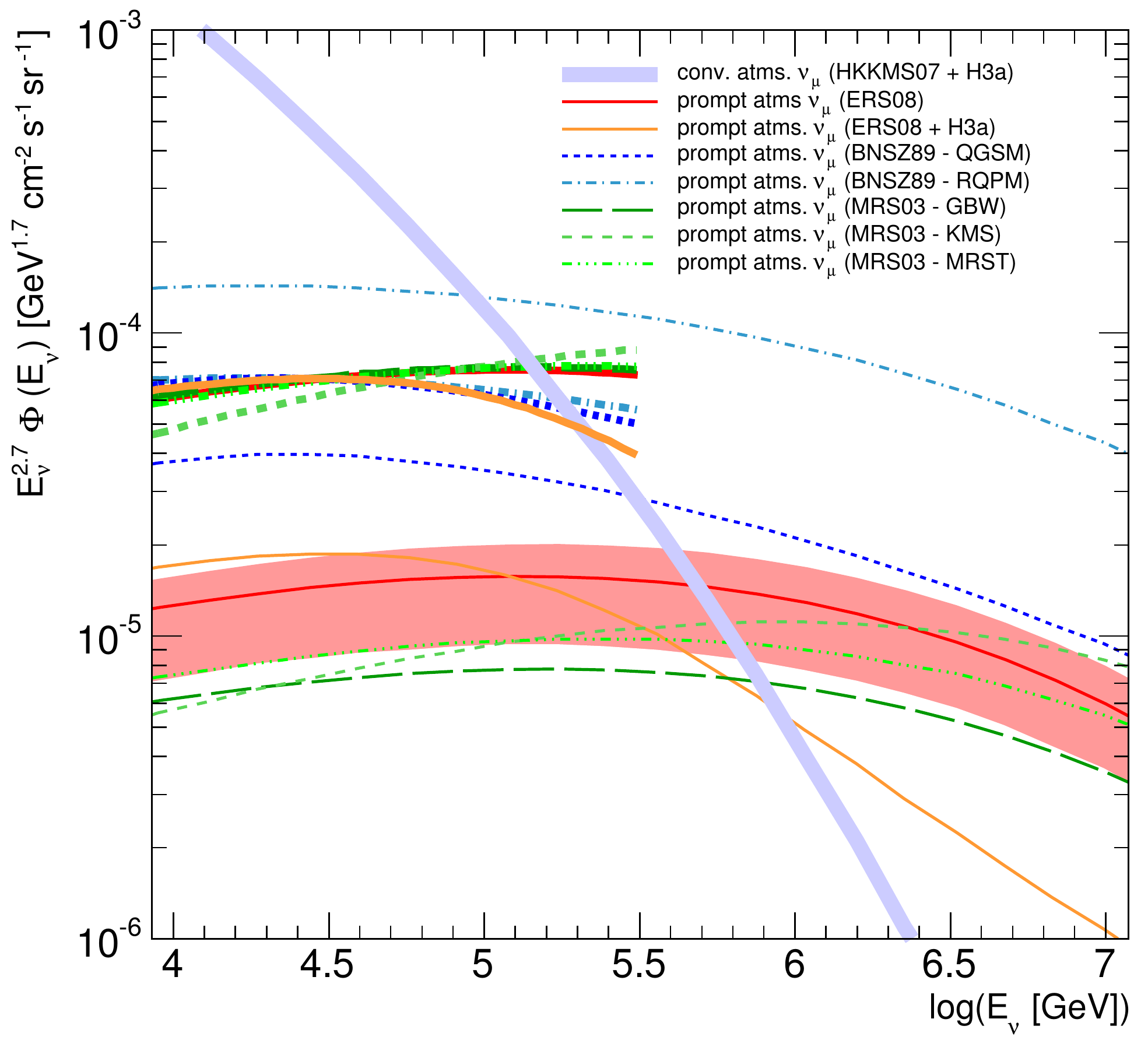}
  \caption{Prompt atmospheric $\nu_{\mu} + \overline \nu_{\mu}$ neutrino fluxes in comparison to the expected flux of conventional atmospheric neutrinos (Honda \textit{et al.}~+ H3a). Model predictions are represented by thin lines \cite{Enberg:2008te, Martin:2003us, Bugaev:1989we}. The red shaped area marks the theoretical uncertainty on the prediction of ERS08. Limits for each model are shown as thick lines in the corresponding line style and color in the valid energy range between $2.3\,$TeV and $360\,$TeV (see Tab.~\ref{tab:PromptMRFs}). The baseline model used here is the model ERS08 modified with the cosmic-ray parameterization by Gaisser \textit{et al.}~and is represented by the thick orange line. Other models are shown as published.}%
  \label{img:PromptLimit}%
\end{figure}

In addition to the limit on the baseline flux ERS08 + H3a, upper limits are also calculated for other prompt neutrino flux predictions \cite{Enberg:2008te, Martin:2003us, Bugaev:1989we}. The results for the different models are given in Tab.~\ref{tab:PromptMRFs} and are shown in Fig.~\ref{img:PromptLimit}. The shape of all prompt models is very similar, with the largest difference in their absolute normalization. For all models, the best fit result for the prompt neutrino component is zero.
The upper limits for additional prompt flux predictions are calculated based on the Wilks' theorem $\chi^2$ approximation to the likelihood ratio distribution and are given in units of model rejection factors (MRF). All limits are valid in the energy range between $2.3\,$TeV and $360\,$TeV.\\

The upper limits derived from this analysis are typically still a factor of $4$ to $10$ above current prompt flux calculations based on perturbative QCD (Enberg \textit{et al.}~\cite{Enberg:2008te}, Martin \textit{et al.}~\cite{Martin:2003us}). The model rejection factor for the intrinsic charm model by Bugaev \textit{et al.}~\cite{Bugaev:1989we} is $0.5$. This means that even a flux as low as $50\%$ of the Bugaev \textit{et al.}~prediction is excluded by this analysis with $90\%$ CL.

\begin{table}
  \footnotesize%
  \caption{Model rejection factors for different theoretical predictions of prompt atmospheric neutrino fluxes \cite{Enberg:2008te, Martin:2003us, Bugaev:1989we}. If not noted otherwise, these models are the original published models and have not been modified for a more accurate cosmic-ray flux parameterization. Except for the baseline model ERS08 with H3a knee, MRFs are based on a $\chi^2$-approximation.}
  \begin{center}
    \begin{tabular}{l|c}
      \hline
        Model & MRF\\
      \hline
	ERS08 + H3a \cite{Enberg:2008te,Gaisser:2012zz} & $3.8$\\
	ERS08 \cite{Enberg:2008te} & $4.8$\\
	ERS08 (max) \cite{Enberg:2008te} & $3.8$\\
	ERS08 (min) \cite{Enberg:2008te} & $8.2$\\
	MRS03 (GBW) \cite{Martin:2003us} & $9.9$\\
	MRS03 (MRST) \cite{Martin:2003us} & $8.0$\\
	MRS03 (KMS) \cite{Martin:2003us} & $8.3$\\
	BNSZ89 (RQPM) \cite{Bugaev:1989we} & $0.5$\\
	BNSZ89 (QGSM) \cite{Bugaev:1989we}& $1.8$\\
      \hline
    \end{tabular}
  \end{center}
  \label{tab:PromptMRFs}
\end{table}

\section{Conclusion}
\label{conclusion}

For the search for a diffuse astrophysical flux of $\nu_\mu + \bar\nu_\mu $ a global likelihood fit of the two-dimensional distribution of measured energy loss and arrival direction of detected muon neutrino events by IceCube in its 59-string configuration was performed. In particular the high neutrino purity of the sample and the careful treatment of systematic uncertainties of the detection method and the theoretical modeling of background in the fit allowed a substantial increase in sensitivity compared to earlier IceCube analyses. With the search presented here, a sensitivity below the Waxman and Bahcall upper bound has been achieved for the first time by a neutrino telescope.

This search found a high-energy excess of $1.8\sigma$ compared to the background scenario of a pure conventional atmospheric model. 
The corresponding best-fit astrophysical $\nu_{\mu} + \overline \nu_{\mu}$ flux
is $E_{\nu}^2 \Phi (E_{\nu}) = 0.25 \cdot 10^{-8}\,\textrm{GeV}\,\textrm{cm}^{-2}\,\textrm{s}^{-1}\,\textrm{sr}^{-1}$. The 90\% confidence level upper limit on the 
flux is $E_{\nu}^2 \Phi (E_{\nu}) \le 1.44 \cdot 10^{-8}\,\textrm{GeV}\,\textrm{cm}^{-2}\,\textrm{s}^{-1}\,\textrm{sr}^{-1}$. 
Due to the observed excess of high-energy events, this limit is slightly above the Waxman and Bahcall upper bound.

This analysis also sets constraints on a prompt atmospheric neutrino flux. The limit on a prompt muon neutrino contribution to the data sample 
is $\Phi_{\textrm{\tiny{prompt}}} (E_{\nu}) \le 3.8 \cdot \Phi_{\textrm{\tiny{Enberg \textit{et al.}~+ H3a}}} (E_{\nu})$. This limit is still a factor of 4 to 10 above pQCD model predictions \cite{Enberg:2008te, Martin:2003us, Bugaev:1989we}, but lowers previous flux constraints by one order of magnitude \cite{Achterberg:2007qp}. The intrinsic charm model by Bugaev \textit{et al.}~is disfavored at a confidence level of more than $90\%$.

%The sensitivity to both, astrophysical and prompt atmospheric neutrinos, can further be enhanced by the addition of cascade event signatures. Also, the combination of different analyses, optmized for different event signatures and energy regimes, is possible.

This result is consistent with the excess of high-energy events found in IceCube analyses searching for cascade-like signatures: An analysis with the IC40 detector found an excess of $2.7\,\sigma$ over the atmospheric background \cite{eike}. The recently reported evidence for an extraterrestrial neutrino flux found in a search with IceCube's IC79 and first year of IC86 configuration has a significance of $4.1\,\sigma$. The upper limits and best-fit fluxes of these analyses are consistent with the IC59 analysis of track-like events presented here within their yet large uncertainties.

Future studies in all detection channels will reveal if the observed excesses can be attributed to an astrophysical neutrino signal. With a runtime of several years, the full IceCube detector will also improve its sensitivity to prompt atmospheric neutrinos and reach the sensitivity level of current prompt model predictions. The muon neutrino search presented here will profit from the larger detector providing higher statistics in particular in the high-energy region. Since high-energy neutrinos are absorbed by the Earth, the effective area for the highest-energy events is largest at the horizon. An extension of the field of view of the muon neutrino search above the horizon would increase the analysis sensitivity to an astrophysical flux. Future muon neutrino searches will reach the sensitivity to probe the astrophysical diffuse neutrino flux at the level of the high-energy starting event analysis reported in Ref.~\cite{hese}.

\begin{acknowledgments}

We acknowledge the support from the following agencies:
U.S. National Science Foundation-Office of Polar Programs,
U.S. National Science Foundation-Physics Division,
University of Wisconsin Alumni Research Foundation,
the Grid Laboratory Of Wisconsin (GLOW) grid infrastructure at the University of Wisconsin - Madison, the Open Science Grid (OSG) grid infrastructure;
U.S. Department of Energy, and National Energy Research Scientific Computing Center,
the Louisiana Optical Network Initiative (LONI) grid computing resources;
Natural Sciences and Engineering Research Council of Canada,
WestGrid and Compute/Calcul Canada;
Swedish Research Council,
Swedish Polar Research Secretariat,
Swedish National Infrastructure for Computing (SNIC),
and Knut and Alice Wallenberg Foundation, Sweden;
German Ministry for Education and Research (BMBF),
Deutsche Forschungsgemeinschaft (DFG),
Helmholtz Alliance for Astroparticle Physics (HAP),
Research Department of Plasmas with Complex Interactions (Bochum), Germany;
Fund for Scientific Research (FNRS-FWO),
FWO Odysseus programme,
Flanders Institute to encourage scientific and technological research in industry (IWT),
Belgian Federal Science Policy Office (Belspo);
University of Oxford, United Kingdom;
Marsden Fund, New Zealand;
Australian Research Council;
Japan Society for Promotion of Science (JSPS);
the Swiss National Science Foundation (SNSF), Switzerland;
National Research Foundation of Korea (NRF);
Danish National Research Foundation, Denmark (DNRF)

\end{acknowledgments}

\begin{appendix}

\section{Calculation of the neutrino knee}
\label{appendix-knee}

The effect of the knee on the neutrino spectrum is estimated relative to the initial power-law extrapolation by calculating the ratio of the neutrino flux at energy $E_\nu$ from the spectrum with the knee to that obtained from the power-law extrapolation. The calculation of Ref.~\cite{Honda:2006qj} is taken as the default spectrum for the power-law extrapolation. The rescaling factor is given by
%\begin{equation}
%\begin{split}
%& \frac{d\Phi_{\nu}{\textrm{\tiny{CR}}}}{dE_{\nu}} / \frac{d\Phi_{\nu}^{\textrm{\tiny{HKKMS07}}}}{dE_{\nu}} \\
%& = \frac{ \int dE_N \frac{d \Phi_{\textrm{\tiny{CR}}}}{dE_N} \frac{d}{dE_{\nu}} Y \left(E_N, E_{\nu}, \cos (\theta^*) \right) }
%{ \int dE_N \frac{d \Phi_{\textrm{\tiny{N,\,Honda}}}}{dE_N} \frac{d}{dE_{\nu}} Y \left(E_N, E_{\nu}, \cos (\theta^*) \right) }
%\end{split}
%\end{equation}
\begin{equation}
\begin{split}
& \frac{\Phi_{\nu}{\textrm{\tiny{CR}}} (E_{\nu})}{\Phi_{\nu}^{\textrm{\tiny{HKKMS07}}} (E_{\nu})} \\
& = \frac{ \int dE_N \Phi_{\textrm{\tiny{N\,CR}}} (E_N) \frac{d}{dE_{\nu}} Y \left(E_N, E_{\nu}, \cos (\theta^*) \right) }
{ \int dE_N \Phi_{\textrm{\tiny{N,\,\textrm{\tiny{HKKMS}}}}} (E_N) \frac{d}{dE_{\nu}} Y \left(E_N, E_{\nu}, \cos (\theta^*) \right) }
\end{split}
\end{equation}
where $\phi_{N,\,\textrm{\tiny{HKKMS}}} (E_N)$ is the spectrum of nucleons used in the calculation of Ref.~\cite{Honda:2006qj} and $\Phi_{\textrm{\tiny{N\,CR}}}$ is the spectrum of nucleons for a different cosmic-ray flux parameterization, here Gaisser H3a \cite{Gaisser:2012zz} or poly-gonato (modified) \cite{hoerandel2003knee}. The yield of $\nu_{\mu} + \bar{\nu}_\mu$ per nucleon as a function of the neutrino energy $E_{\nu}$, the nucleon energy $E_N$ and the inclination of the cosmic ray $\theta^*$ is taken as
\begin{equation}
\begin{split}
& Y \left(E_N, E_{\nu}, \cos (\theta^*) \right) = \\
& \frac{\epsilon_{\nu}^*}{E_{\nu} \cos (\theta^*)} \cdot \left( \frac{E_N}{A E_{\nu}} \right)^p \left( 1 - \frac{A E_{\nu}}{E_N}\right)^q
\end{split}
\label{eq:elbert}
\end{equation}
where $\epsilon_\nu^*\,=\,4.8$~GeV, $p=0.76$ and $q = 5.25$. 

\begin{figure}[h]
  \includegraphics[width=\linewidth]{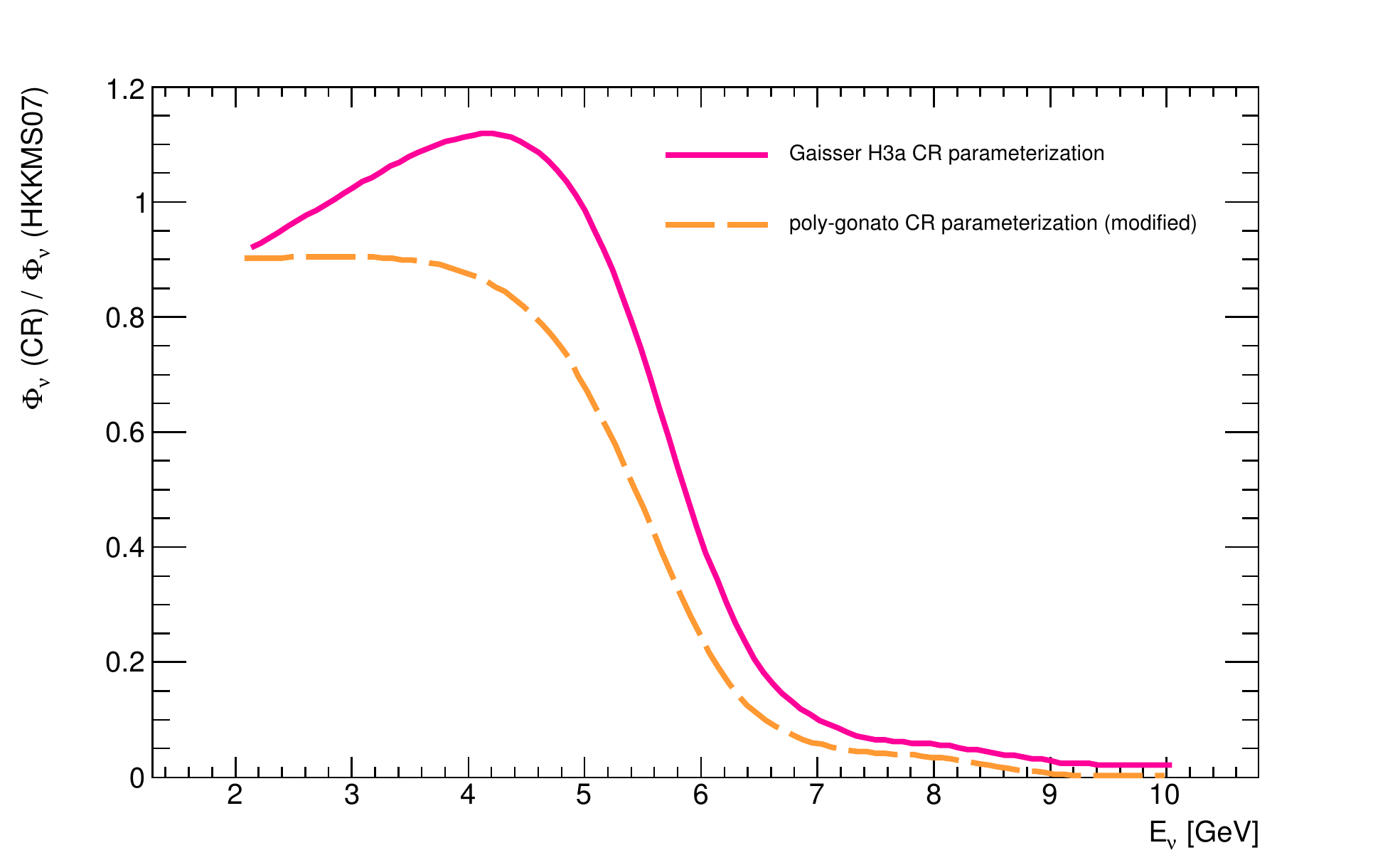}
  \caption{Ratio of the two muon neutrino fluxes calculated based on the cosmic-ray parameterizations H3a \cite{Gaisser:2012zz} and poly-gonato (modified) \cite{hoerandel2003knee} with knee to the standard HKKMS07 muon neutrino flux \cite{Honda:2006qj} as a function of energy.}%
  \label{img:neutrinofluxeswithknee}
\end{figure}

\vspace{30pt}

Equation~\ref{eq:elbert} is an adaptation of the approximation originally proposed by Elbert~\cite{Elbert} to approximate the number of muons per primary nucleon. It is based on air shower simulations (see e.g. Refs.~\cite{GS, Forti}). 
It has been checked that the formula for neutrinos is consistent with simulations with CORSIKA and SIBYLL over a range of primary energies.
This rescaling method is used to be able to take advantage of the full range of existing simulations of atmospheric neutrino-induced muons in IceCube. The rescaling factors are shown in Fig.~\ref{img:neutrinofluxeswithknee}.

Similar correction functions have also been calculated for the prompt atmospheric neutrino prediction ERS08~\cite{Enberg:2008te}. Since the analytical derivation of a neutrino yield factor is challenging for prompt neutrinos, the yield is calculated from air shower simulations using CORSIKA \cite{Heck:1998vt} with DPMJET \cite{Battistoni:1995yv, Berghaus:2007hp}.

\end{appendix}

\bibliography{references}

\end{document}